\documentclass[a4paper,11pt]{article}

\usepackage[dvipsnames, svgnames, x11names]{xcolor}

\usepackage[numbers,sort&compress]{natbib}
\usepackage[all]{xy}
\usepackage{amsmath,amsfonts,amssymb,amsthm,epsfig,amscd,comment,latexsym,psfrag}
\usepackage{CJK}
\usepackage{mathrsfs}
\usepackage{nicefrac,xspace,tikz}
\usepackage{arydshln}
\usetikzlibrary{arrows}
\usepackage{xcolor,float}
\usepackage{appendix}
\usepackage{txfonts}
\usepackage{pfnote}
\makeatletter

\newcommand{\Rmnum}[1]{\expandafter\@slowromancap\romannumeral #1@}
\makeatother

\def\footnoterule{\kern 1mm \hrule width 7cm \kern 2.2mm}%
\renewcommand{\thefootnote}{\fnsymbol{footnote}}

\def\dsum{\displaystyle\sum}

\def\Tr{\mathrm{Tr}}

\def\Ker{\mathrm{Ker}}
\def\span{\mathrm{span}}
\def\Coker{\mathrm{Coker}}
\def\level{\mathrm{level}}

\numberwithin{equation}{section}

\renewcommand{\a}{\alpha}
\renewcommand{\b}{\beta}

\usepackage{graphicx}
\usepackage{float}
\usepackage{subfigure}
\usepackage[]{caption2}

\renewcommand{\thesubfigure}{(\roman{subfigure})}
\makeatletter \renewcommand{\@thesubfigure}{\thesubfigure \space}
\renewcommand{\p@subfigure}{} \makeatother

\renewcommand\appendix{\par
    \setcounter{section}{0}
    \setcounter{subsection}{0}
    \gdef\thesection{Appendix \Alph{section}}}

\allowdisplaybreaks
\begin{document}
\begin{titlepage}
\begin{center}
{\Large\bf A two-tensor model with order-three}\vskip .2in
{\large Bei Kang$^{a}$,\footnote{kangbei@ncwu.edu.cn}
Lu-Yao Wang$^{b}$,\footnote{wangly100@outlook.com}  Ke Wu$^{b}$\footnote{wuke@cnu.edu.cn}
and Wei-Zhong Zhao$^{b,}$\footnote{Corresponding author: zhaowz@cnu.edu.cn}} \vskip .2in
$^a${\em School of Mathematics and Statistics, North China University of Water Resources and Electric Power,
Zhengzhou 450046, Henan, China}\\
$^b${\em School of Mathematical Sciences, Capital Normal University,
Beijing 100048, China} \\

\begin{abstract}
We construct a two-tensor model with order-3 and present its $W$-representation.
Moreover we derive the compact expressions of correlators from the $W$-representation and analyze
the free energy in large $N$ limit. In addition, we establish the correspondence between two
colored Dyck walks in the Fredkin spin chain and tree operators in the ring. Based on the
classification Dyck walks, we give the number of tree operators with the given level. Furthermore,
we show the entanglement scaling of Fredkin spin chain beyond logarithmic scaling in the ordinary
critical systems from the viewpoint of tensor model.
\end{abstract}

\end{center}

{\small Keywords: Matrix Models, Conformal and $W$ Symmetry}

\end{titlepage}

\section{Introduction}
Matrix models can be associated with discretized random surfaces and $2D$ quantum gravity \cite{Francesco1995}.
For the perturbative series of matrix models, one can rewrite them as a series in $1/N$ indexed by
the genus, where $N$ is the size of the matrix. At leading order, the ordinary topological expansion
of matrix models dominated by planar graphs. This limit has been widely studied and has led to a
plethora of results, in particular their continuum limit and integrability. As the generalizations
of matrix models from matrices to tensors, tensor models were originally introduced to describe the
higher dimensional quantum gravity \cite{Jonsson}-\cite{Sasakura}. The $1/N$ expansion for colored
tensor models was identified in \cite{Gurau2011}-\cite{GurauRyan2012}, where colored tensors
are tensors with no further tensorial symmetry assumed (and an additional flavor index called
color). It was shown that tensor models admit a specific kind of large $N$ limit, the so-called
melonic limit \cite{Gurau2011}-\cite{Gurau2016}.
In this limit, the dominant melonic graphs are built recursively by two-point insertions on an
initial two-point diagram. Much progress has been made in recent years on colored tensor models,
which has stimulated other fields. For instance, the existence of $1/N$ expansion for tensor has
triggered a series of investigations on the renormalization group analysis of the quantum field
theoretic counterpart of tensor models, i.e. tensor field theory \cite{Oriti}-\cite{Benedetti}.
Importantly, it was shown that in specific instances, these quantum field models are asymptotically
free \cite{Geloun2012}-\cite{Rivasseau2015}. Concerning the large $N$ expansion of tensor models, it
turns out that coloration was not a prerequisite for the discovery of such an expansion : Gurau
\cite{Gurau2018} proposed a new approach to the $1/N$ expansion adapted to symmetric (and
antisymmetric) tensors. It was proved that a tensor model with two symmetric tensors and
interactions the complete graph admits a $1/N$ expansion. Moreover, for the symmetric traceless and
the antisymmetric tensor models in rank-3 with tetrahedral interaction, it was found that they admit
a $1/N$ expansion \cite{Kolanowski2019}. Universally, at leading order, all these models are dominated
by melonic diagrams.

Tensor models have found also intriguing connections with gauge-string duality. By applying
permutation TFT methods, Ben Geloun and Ramgoolam  \cite{jbgeloun1} counted gauge invariants for
tensor models and obtained some formulae for correlators of the tensor model invariants.
Furthermore, they showed that the counting of observables and correlators for a $3$-index tensor
model are organized by the structure of a family of permutation centralizer algebras
\cite{jbgeloun2}. Collective field theory provides a systematic construction of the dynamics of
invariant observables of the theory. It has been applied in the analysis of tensor models, such as
the large $N$ dynamics of the boson tensor model quantum mechanics \cite{Tribelhorn} and holographic
duals of tensor models \cite{Mahu2020}. Rainbow tensor model which has gauge symmetry
$U(N_1)\bigotimes \cdots \bigotimes U(N_r)$ is a direct generalization
of rectangular complex matrix model with rectangular matrix substituted by a complex-valued tensor
of rank-$r$ \cite{ItoyamaJHEP2017}-\cite{ItoyamaJHEP2019}. In rainbow tensor model, all the planar
diagrams are automatically melonic. Once again, just as any tensor models with no symmetry assumed
between the tensor indices, melonic graphs are dominant in the large $N$ limit of rainbow tensor
models. The simplest rainbow tensor model is the Aristotelian RGB (red-green-blue) model with a
single complex tensor of rank-3 and the RGB symmetry $U({\textcolor{red}{N_1}})\bigotimes
U({\textcolor{ForestGreen}{N_2}})\bigotimes U({\textcolor{blue}{N_3}})$.
With the example of the Aristotelian RGB model, Itoyama et al. \cite{ItoyamaJHEP2017} introduced a
few methods which allow one to connect calculations in the tensor models to those in the matrix
models. Furthermore, it was found that there are some new factorization formulas and sum rules
for the Gaussian correlators in the Hermitian and complex matrix theories, square and rectangular.

$W$-representation of matrix model was proposed by Morozov and Shakirov for the realization of
partition function by acting on elementary functions with exponents of the given $W$-operator \cite{Shakirov2009}.
The superintegrable matrix models can be analyzed from the viewpoint of $W$-representations
\cite{Wangr2022}-\cite{MMMP}. The progress has been made on $W$-representations of tensor models.
As was expected, the Gaussian tensor model with Gaussian action can be realized by the
$W$-representation \cite{Itoyama2020}. Moreover, from its character expansion with respect to the
Schur functions, it turned out that this model is superintegrable. The $W$-representation of rainbow
tensor model has been investigated \cite{BKang2021, LYWang2021}. In this paper, we will make a step
towards the $W$-representation of a two-tensor model with
order-3\renewcommand{\thefootnote}{\arabic{footnote}}
\footnote{Here we use ``order" instead of ``rank" since the rank of a tensor is defined by the minimal
number of tensors of order 1 required to express a tensor as a sum of such tensors.}$^,
$ \footnote{The two-tensor model that we deal with can be seen as the two-tensor extension  associated
with ``uncolored" tensor model defined in \cite{Bonzom}.}, and analyze the correlators and free energy.

This paper is organized as follows.
In section 2, we give the keystone operators and construct a graded ring with tree and loop
operators. The kernel and cokernel of the cut-and-join structure in the graded ring are
investigated. Then we establish the correspondence between two colored Dyck walks in the Fredkin
spin chain and tree operators. Based on the classification Dyck walks, we present the number of tree
operators with the given level. Moreover for the entanglement entropy of the Fredkin spin chain, we
show the entanglement scaling beyond logarithmic scaling in the ordinary critical systems from the
viewpoint of tensor model.
In section 3, we construct a two-tensor model with order-3 and present its $W$-representation.
In terms of the $W$-representation, we derive the compact expression of correlators. Using the
collective field, we also calculate the correlators. Furthermore we discuss the free energy and its large $N$ limit.
We end this paper with the conclusions in section 4.
We list some results of cut operation $\Delta$ and kernel in the Appendix.

\section{The graded ring of gauge-invariant operators }

\subsection{Gauge invariants and cut-join operators}
Let us consider the tensor fields~$A_i^{{j_1},{j_2}}$ and $B_i^{{j_1},{j_2}}$ with order-3 which
transform in the fundamental of~$G=U(N_1)\times U(N_2)\times U(N_3)$,
where~$i=1,\cdots,N_1$,~$j_1=1,\cdots,N_2$ and~$j_2=1,\cdots,N_3$.
We denote $V_k$ the vector space carrying a copy of the fundamental representation of~$U(N_k)$.
The fields transforming in the fundamental of~$G=U({N_1})\times U({N_2})\times U({N_3})$
belong to~$V_1\times V_2\times V_3$.
To build gauge invariants, we introduce the conjugate tensor fields that transform in the anti-fundamental,
denoted~$\bar{A}^{{i}}_{{{j_1}},{{j_2}}}$ and~$\bar{B}^{i}_{{j_1},{j_2}}$.
Then the gauge invariants with level-$(n+m)$ are given by contracting corresponding upper and lower
indices, denoted formally as \cite{jbgeloun1,Tribelhorn}
\begin{eqnarray}\label{tsigma}
\mathcal{T}_{\check{\sigma}}^{(n,m)}
\equiv\mathcal{T}_{(\sigma_1,\sigma_2,\sigma_3)}^{(n,m)}
=\mathcal{A}_{I}^{J_1,J_2}\cdot(\sigma_1,\sigma_2,\sigma_3)\cdot\bar{\mathcal{A}}^{I}_{J_1,J_2}
=\mathcal{A}_{I}^{J_1,J_2}\bar{\mathcal{A}}^{\sigma_1(I)}_{\sigma_2(J_1),\sigma_3(J_2)},
\end{eqnarray}
where~$n$ and $m$ are non-negative integers satisfying~$n+m\geqslant1$, and the action
$\check{\sigma}\equiv(\sigma_1,\sigma_2,\sigma_3)\in S_{n+m}\times S_{n+m}\times H_{n,m}$ acts
independently on three indices of~$\bar{\mathcal{A}}^{I}_{J_1,J_2}$,
where $S_{n+m}$ is the permutation group, and~$H_{n,m}\equiv S_n\times S_m$ is a subgroup of~$S_{n+m}$.
The sleek notation uses the capital Roman letters $I,\ J_1,\ J_2$ to collect all of the little Roman
letter indices, for example $I$ stands for $i^{(1)},i^{(2)},\cdots,i^{(n+m)}$.
Repeated indices are summed over in (\ref{tsigma}) and the following text.
~$\mathcal{A}_{I}^{J_1,J_2}$,~$\bar{\mathcal{A}}^{I}_{J_1,J_2}$
and~$\bar{\mathcal{A}}^{\sigma_1(I)}_{\sigma_2(J_1),\sigma_3(J_2)}$ are given by
\begin{eqnarray}
\mathcal{A}_{I}^{J_1,J_2}&=&A_{i^{(1)}}^{j^{(1)}_1,j^{(1)}_{2}}\cdots A_{i^{(n)}}^{j^{(n)}_1,j^{(n)}_{2}}
B_{i^{(n+1)}}^{j^{(n+1)}_1,j^{(n+1)}_{2}}\cdots B_{i^{(n+m)}}^{j^{(n+m)}_1,j^{(n+m)}_{2}},\nonumber\\
\bar{\mathcal{A}}^{I}_{J_1,J_2}&=&\bar A^{i^{(1)}}_{j^{(1)}_1,j^{(1)}_{2}}\cdots \bar A^{i^{(n)}}_{j^{(n)}_1,j^{(n)}_{2}}
\bar B^{i^{(n+1)}}_{j^{(n+1)}_1,j^{(n+1)}_{2}}\cdots \bar B^{i^{(n+m)}}_{j^{(n+m)}_1,j^{(n+m)}_{2}},\nonumber\\
\bar{\mathcal{A}}^{\sigma_1(I)}_{\sigma_2(J_1),\sigma_3(J_2)}
&=&(\sigma_1,\sigma_2,\sigma_3)\cdot\bar{\mathcal{A}}^{I}_{J_1,J_2}\nonumber\\
&=&
\bar A_{j^{\sigma_2(1)}_1,j^{\sigma_3(1)}_{2}}^{i^{\sigma_1(1)}}\cdots\bar A_{j^{\sigma_2(n)}_1,j^{\sigma_3(n)}_{2}}^{i^{\sigma_1(n)}}
\bar B_{j^{\sigma_2(n+1)}_1,j^{\sigma_3(n+1)}_{2}}^{i^{\sigma_1(n+1)}}
\cdots \bar B_{j^{\sigma_2(n+m)}_1,j^{\sigma_3(n+m)}_{2}}^{i^{\sigma_1(n+m)}}.
\end{eqnarray}
Here we denote $\sigma_1(I)=(i^{\sigma_1(1)},\cdots,i^{\sigma_1(n+m)})$ and similarly for $\sigma_2(J_1)$ and $\sigma_3(J_2)$.
Note that~$\mathcal{T}_{\check{\sigma}}^{(0,m)}$ and~$\mathcal{T}_{\check{\sigma}}^{(n,0)}$ are
exactly the gauge invariants in Aristotelian RGB model \cite{ItoyamaJHEP2017}.
Since~$\sigma_3\in S_{n}\times  S_{m}$, we actually focus on a subset of gauge
invariants with all possible contractions where
\begin{description}
\item[(i)]
for $A_{{i}}^{{{j_1}},{{j_2}}}$ and~$\bar{B}^{{i'}}_{{{j_1'}},{{j_2'}}}$ (or
$B_{{i'}}^{{{j_1'}},{{j_2'}}}$ and ~$\bar{A}^{{i}}_{{{j_1}},{{j_2}}}$ ), we do not contract the indices $j_2$ and $j_2'$;
\item[(ii)]
no contractions are possible for $A_{{i}}^{{{j_1}},{{j_2}}}$ and $B_{{i'}}^{{{j_1'}},{{j_2'}}}$ (and
$\bar{A}^{{i}}_{{{j_1}},{{j_2}}}$ and $\bar{B}^{{i'}}_{{{j_1'}},{{j_2'}}}$).
\end{description}

As done in \cite{ItoyamaJHEP2017}, let us take the following six operators as the so-called keystone
operators
\begin{eqnarray}\label{keystoneope}
&&\mathcal{T}_{((12),id,id)}^{(2,0)}
=A_{{i^{(1)}}}^{{{j_1^{(1)}}},{{j_2^{(1)}}}}
\bar{A}^{{i^{(2)}}}_{{{j_1^{(1)}}},{{j_2^{(1)}}}}
A_{{i^{(2)}}}^{{{j_1^{(2)}}},{{j_2^{(2)}}}}
\bar{A}^{{i^{(1)}}}_{{{j_1^{(2)}}},{{j_2^{(2)}}}},\nonumber\\
&&
\mathcal{T}_{(id,(12),id)}^{(2,0)}
=A_{{i^{(1)}}}^{{{j_1^{(1)}}},{{j_2^{(1)}}}}
\bar{A}^{{i^{(1)}}}_{{{j_1^{(2)}}},{{j_2^{(1)}}}}
A_{{i^{(2)}}}^{{{j_1^{(2)}}},{{j_2^{(2)}}}}
\bar{A}^{{i^{(2)}}}_{{{j_1^{(1)}}},{{j_2^{(2)}}}}
,\nonumber\\
&&
\mathcal{T}_{((12),id,id)}^{(0,2)}
=B_{{i^{(1)}}}^{{{j_1^{(1)}}},{{j_2^{(1)}}}}
\bar{B}^{{i^{(2)}}}_{{{j_1^{(1)}}},{{j_2^{(1)}}}}
B_{{i^{(2)}}}^{{{j_1^{(2)}}},{{j_2^{(2)}}}}
\bar{B}^{{i^{(1)}}}_{{{j_1^{(2)}}},{{j_2^{(2)}}}}
,\nonumber\\
&&
\mathcal{T}_{(id,(12),id)}^{(0,2)}
=B_{{i^{(1)}}}^{{{j_1^{(1)}}},{ {j_2^{(1)}}}}
\bar{B}^{{i^{(1)}}}_{{{j_1^{(2)}}},{ {j_2^{(1)}}}}
B_{ {i^{(2)}}}^{{ {j_1^{(2)}}},{ {j_2^{(2)}}}}
\bar{B}^{ {i^{(2)}}}_{{ {j_1^{(1)}}},{ {j_2^{(2)}}}}
,\nonumber\\
&&
\mathcal{T}_{((12),id,id)}^{(1,1)}
=A_{ {i^{(1)}}}^{{ {j_1^{(1)}}},{ {j_2^{(1)}}}}
\bar{A}^{ {i^{(2)}}}_{{ {j_1^{(1)}}},{ {j_2^{(1)}}}}
B_{ {i^{(2)}}}^{{ {j_1^{(2)}}},{ {j_2^{(2)}}}}
\bar{B}^{ {i^{(1)}}}_{{ {j_1^{(2)}}},{ {j_2^{(2)}}}}
,\nonumber\\
&&
\mathcal{T}_{(id,(12),id)}^{(1,1)}
=A_{ {i^{(1)}}}^{{ {j_1^{(1)}}},{ {j_2^{(1)}}}}
\bar{A}^{ {i^{(1)}}}_{{ {j_1^{(2)}}},{ {j_2^{(1)}}}}
B_{ {i^{(2)}}}^{{ {j_1^{(2)}}},{ {j_2^{(2)}}}}
\bar{B}^{ {i^{(2)}}}_{{ {j_1^{(1)}}},{ {j_2^{(2)}}}}.
\end{eqnarray}
Since $\mathcal{T}_{\check{\sigma}}^{(n,m)}$ with the different $\sigma_1$, $\sigma_2$ and
$\sigma_3$ may give the same gauge invariant, we need not discuss the case of a non $id$ permutation
on the third slot.

For the gauge-invariant operator $\mathcal{T}_{\check{\alpha}}^{(a,b)}$, $\check{\alpha}$ is a
permutation about $a+b$ elements, and $a+b$ is the level of $\mathcal{T}_{\check{\alpha}}^{(a,b)}$,
we denote $\level (\check{\alpha})=a+b$. Let us introduce the cut and join operations on the
gauge-invariant operator $\mathcal{T}_{\check{\alpha}}^{(a,b)}$ as follows.
The actions of the cut operations on the gauge-invariant operator
$\mathcal{T}_{\check{\alpha}}^{(a,b)}$ are
\begin{eqnarray}\label{cut}
\Delta \mathcal{T}_{\check{\alpha}}^{(a,b)}&=&
\sum_{i=1}^{ {N_{1}}}
\sum_{j_1=1}^{ {N_{2}} }
\sum_{j_{2}=1}^{  {N_{3}}}
\dfrac{\partial^2 \mathcal{T}_{\check{\alpha}}^{(a,b)}}{\partial A_i^{j_1,j_{2}}
\partial \bar{A}^i_{j_1,j_{2}}}\nonumber\\
&=&
\sum_{k=1}^3\sum_{\substack{a_1+\cdots+a_k+1=a\\ b_1+\cdots+b_k=b\\a_1\leq a_2\leq\cdots\leq a_k}}\sum_{\check{\beta}_1,\cdots,\check{\beta}_k}
{\Delta}_{\check{\alpha},\check{\beta}_1,\cdots,\check{\beta}_k}^{(a,b),(a _1,b_1),\cdots,(a_k,b_k)}
\mathcal{T}_{\check{\beta}_1}^{(a_1,b_1)}\cdots\mathcal{T}_{\check{\beta}_k}^{(a_k,b_k)},\ a+b\geqslant 2,
\nonumber\\
\tilde{\Delta} \mathcal{T}_{\check{\alpha}}^{(a,b)}
&=&\sum_{i=1}^{ {N_{1}}}\sum_{j_1=1}^{ {N_{2}} }\sum_{j_{2}=1}^{  {N_{3}}}
\dfrac{\partial^2 \mathcal{T}_{\check{\alpha}}^{(a,b)}}{\partial B_i^{j_1,j_{2}}
\partial \bar{B}^i_{j_1,j_{2}}}\nonumber\\
&=&\sum_{k=1}^3\sum_{\substack{a_1+\cdots+a_k=a\\ b_1+\cdots+b_k+1=b\\b_1\leq b_2\leq\cdots\leq b_k}}
\sum_{\gamma_1,\cdots,\gamma_k}
{\tilde{\Delta}}_{\check{\alpha},\check{\gamma}_1,\cdots,\check{\gamma}_k}^{(a,b),(a _1,b_1),\cdots,(a_k,b_k)}
\mathcal{T}_{\check{\gamma}_1}^{(a_1,b_1)}\cdots\mathcal{T}_{\check{\gamma}_k}^{(a_k,b_k)},\ a+b\geqslant 2,
\end{eqnarray}
where  $\level (\check{\beta}_i)=a_i+b_i\ (i=1,\cdots,k)$, $\level (\check{\gamma}_j)=a_j+b_j\ (j=1,\cdots,k)$,
${\Delta}_{\check{\alpha},\check{\beta}_1,\cdots,\check{\beta}_k}^{(a,b),(a _1,b_1),\cdots,(a_k,b_k)}$ and
${\tilde{\Delta}}_{\check{\alpha},\check{\gamma}_1,\cdots,\check{\gamma}_k}^{(a,b),(a _1,b_1),\cdots,(a_k,b_k)}$
are polynomials in $N_i$ with integer coefficients (see examples in (\ref{cutlevel})).

The  actions of the join operations on the gauge-invariant operator
$\mathcal{T}_{\check{\alpha}}^{(a,b)}$ and $\mathcal{T}_{\check{\beta}}^{(c,d)}$ are
\begin{eqnarray}\label{join}
\{\mathcal{T}_{\check{\alpha}}^{(a,b)},\mathcal{T}_{\check{\beta}}^{(c,d)}\}_A&=&
\sum_{i=1}^{ {N_{1}}}\sum_{j_1=1}^{ {N_{2}}}
\sum_{j_{2}=1}^{ {N_{3}}}
\dfrac{\partial \mathcal{T}_{\check{\alpha}}^{(a,b)}}{\partial A_i^{j_1,j_{2}}}
\dfrac{\partial \mathcal{T}_{\check{\beta}}^{(c,d)}}{ \partial \bar{A}^i_{j_1,j_{2}}}\nonumber\\
&=&
\sum_{\check{\gamma}\ |\ \level(\check{\gamma})=a+b+c+d-1}
{\Lambda}_{\check{\alpha},\check{\beta},\check{\gamma}}^{(a,b),(c,d),(a+c-1,b+d)}
\mathcal{T}_{\check{\gamma}}^{(a+c-1,b+d)},
\nonumber\\
\{\mathcal{T}_{\check{\alpha}}^{(a,b)},\mathcal{T}_{\check{\beta}}^{(c,d)}\}_B&=&
\sum_{i=1}^{ {N_{1}}}
\sum_{j_1=1}^{ {N_{2}}}
\sum_{j_{2}=1}^{ {N_{3}}}
\dfrac{\partial \mathcal{T}_{\check{\alpha}}^{(a,b)}}{\partial B_i^{j_1,j_{2}}}
\dfrac{\partial \mathcal{T}_{\check{\beta}}^{(c,d)}}{ \partial \bar{B}^i_{j_1,j_{2}}}\nonumber\\
&=&
\sum_{\check{\sigma}\ |\ \level(\check{\sigma})=a+b+c+d-1}
{\tilde{\Lambda}}_{\check{\alpha},\check{\beta},\check{\sigma}}^{(a,b),(c,d),(a+c-1,b+d)}
\mathcal{T}_{\check{\sigma}}^{(a+c-1,b+d)},
\end{eqnarray}
where~${\Lambda}_{\check{\alpha},\check{\beta},\check{\gamma}}^{(a,b),(c,d),(a+c-1,b+d)}$
and ~${\tilde{\Lambda}}_{\check{\alpha},\check{\beta},\check{\sigma}}^{(a,b),(c,d),(a+c-1,b+d)}$
are integer coefficients (see examples in Table 1).
Here for clarity of classifying and generating structure of the gauge invariants,
the cut and join operations (\ref{cut}) and (\ref{join}) involve only~$A$ and~$\bar{A}$, or~$B$ and~$\bar{B}$,
and do not contain~$A$ and~$\bar{B}$, or~$B$ and~$\bar{A}$.

\begin{figure}[H]
\centering
\includegraphics[height=3cm]{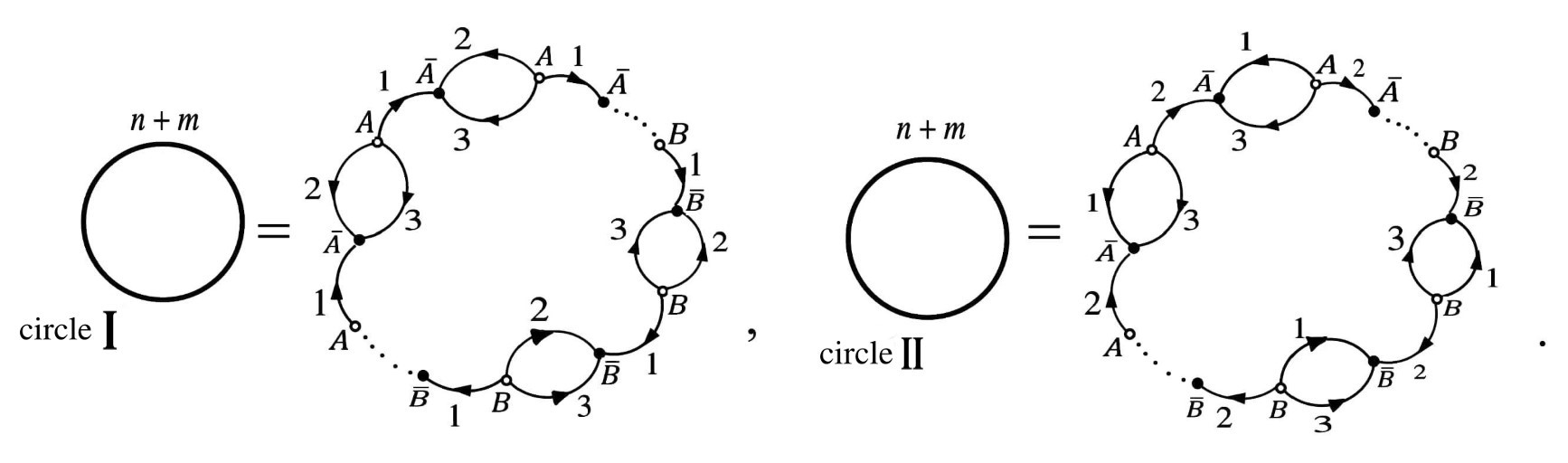}
\caption{ Circles ~\textbf{I}~ and ~\textbf{II}.}
\label{circles}
\end{figure}

To draw the graph associated to the gauge invariants, we represent every tensor field by a white
vertex and its conjugation  by a black vertex. We promote the position of an index to a color: $i$
has color $1$, $j_1$ has color $2$ and $j_2$ has color $3$.  Lines inherit the color of the index,
and always connect a black and a white vertex.
The direction of arrow depends on the choice of covariant and contravariant indices, which is from
the tensor field to its conjugate. The gauge invariants $\mathcal{T}^{(n,m)}_{((12\cdots m), id,
id)}$ and $\mathcal{T}^{(n,m)}_{(id, (12\cdots m), id)}$ with given length-$(n+m)$  can be depicted
as $(n+m)$ indexed circles ``~\textbf{I}~" and ``~\textbf{II}~", respectively (see
Fig.\ref{circles}).

By means of the keystone operators (\ref{keystoneope}), we may construct a graded ring $\mathcal{S}$
with tree and loop operators,
where
$\mathcal{S}$ is generated by the keystone operators with addition, multiplication, cut and join operations.
$\mathcal{S}=\sum_{l=1}^{\infty}\mathcal{S}_{l}$, where~$\mathcal{S}_{l}$ consists of the gauge invariants with level~$l$.
Note that for the tree and loop operators, they have the similar construction rules with the Aristotelian tensor model.
We use the  black dotted lines to represent the Feynman propagators.

\begin{figure}[H]
\centering
\includegraphics[width=0.7\textwidth]{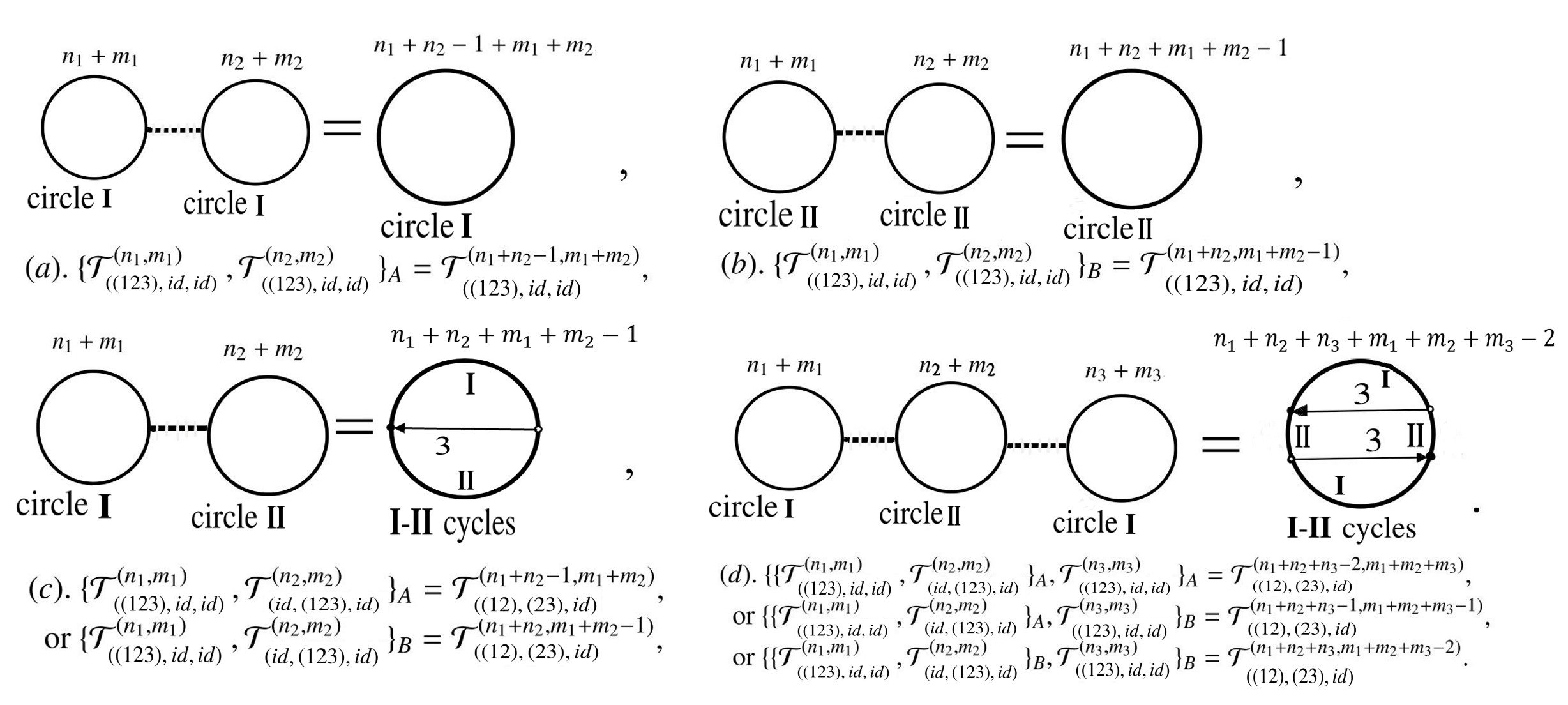}
\caption{Tree operators.}
\label{alltree}
\end{figure}

The tree operators made from $(\ref{keystoneope})$ alone are constructed by merging two vertices in two concentric circles
(propagators) of the same color (see Figs. \ref{alltree}(a) and \ref{alltree}(b)). When tree operators involve chains with both
circles  \textbf{I} and  \textbf{II}, they can be constructed by merging two vertices of two circles (propagators)
of different index (see Figs. \ref{alltree}(c) and \ref{alltree}(d)).

\begin{figure}[H]
\centering
\includegraphics[width=0.7\textwidth]{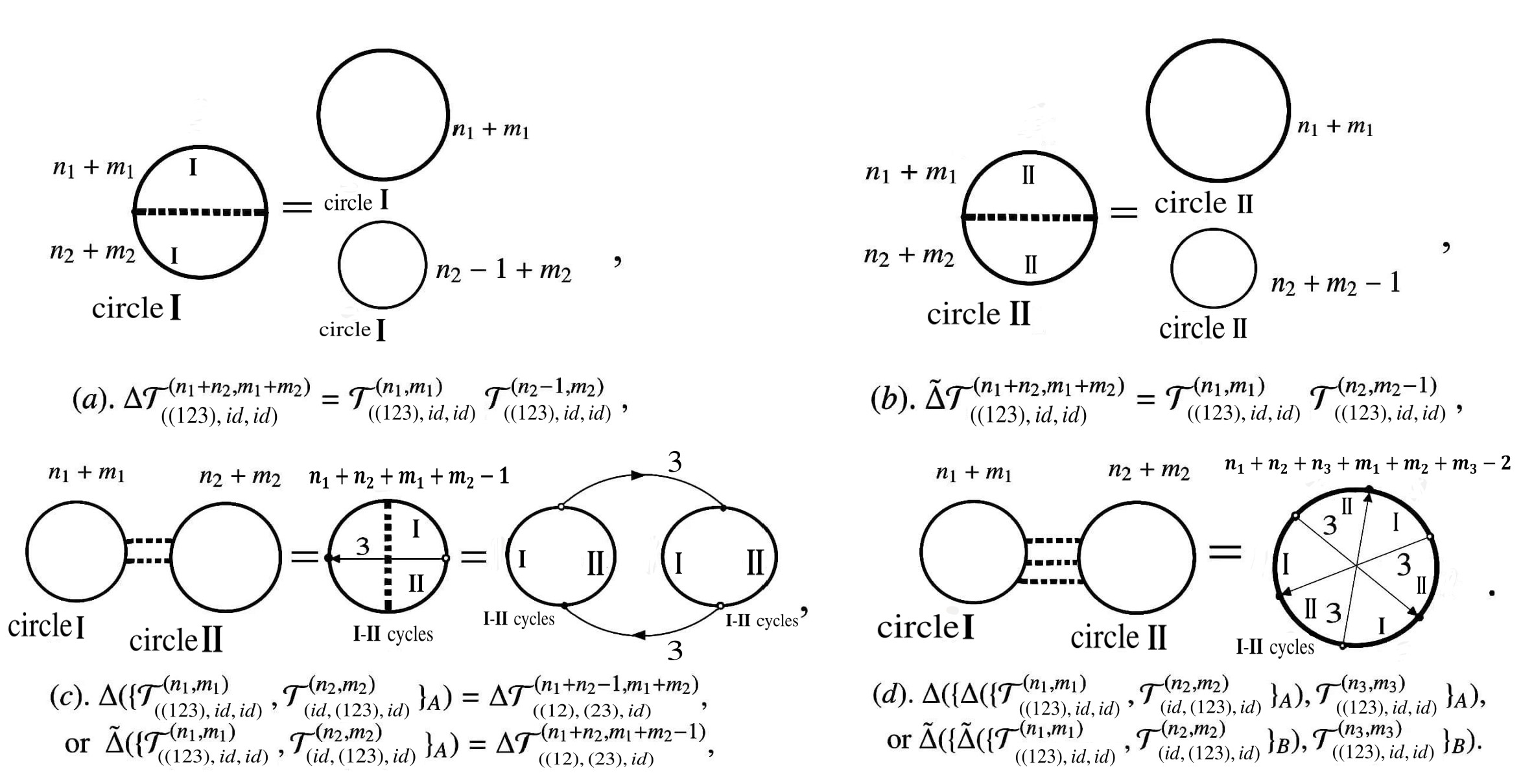}
\caption{Loop operators.}
\label{loopr=3}
\end{figure}

\begin{figure}[H]
\centering
\includegraphics[height=2.55cm]{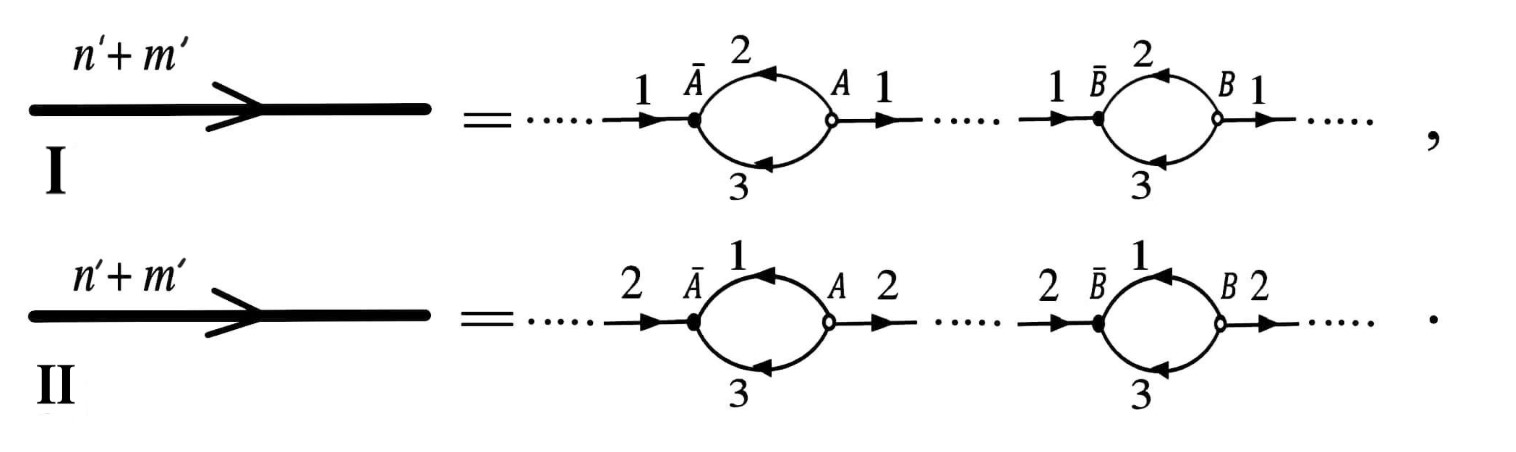}
\caption{Propagators.}
\label{propagators}
\end{figure}

The loop operators made from $(\ref{keystoneope})$ alone are constructed by merging two vertices inside circles (propagators)
(see Figs. \ref{loopr=3}(a) and \ref{loopr=3}(b)).
When loop operators involve both  circles \textbf{I} and \textbf{II}, they are either the \textbf{I}-\textbf{II} cycles with the
intersecting shortcuts of color $3$ or several such \textbf{I}-\textbf{II} cycles with the shortcuts connected by lines of color $3$
(see Figs. \ref{loopr=3}(c) and \ref{loopr=3}(d)), which are constructed by merging two vertices in two circles (propagators) of two different indexes.
Note that \textbf{I}-\textbf{II} cycles form a circle with length-$(n+m)$  divided into segments through the arrow line of color
$3$ (see Figs. \ref{alltree}(c), (d) and Figs. \ref{loopr=3}(c), (d)).
The segments can be drawn as lines labeled ``\textbf{I}" or ``\textbf{II}" with length-$(n'+m')$, and $n'\leq n, m'\leq m$ (see Fig. \ref{propagators}).
As done in Ref.\cite{ItoyamaJHEP2017}, if we consider merging two vertices in two
segments $\textbf{I}$ and $\textbf{II}$, then it leads to emerging of two new inter-propagator vertices connected by a arrow line of color $3$. Pictorially:
\begin{eqnarray}
\includegraphics[height=4.5cm]{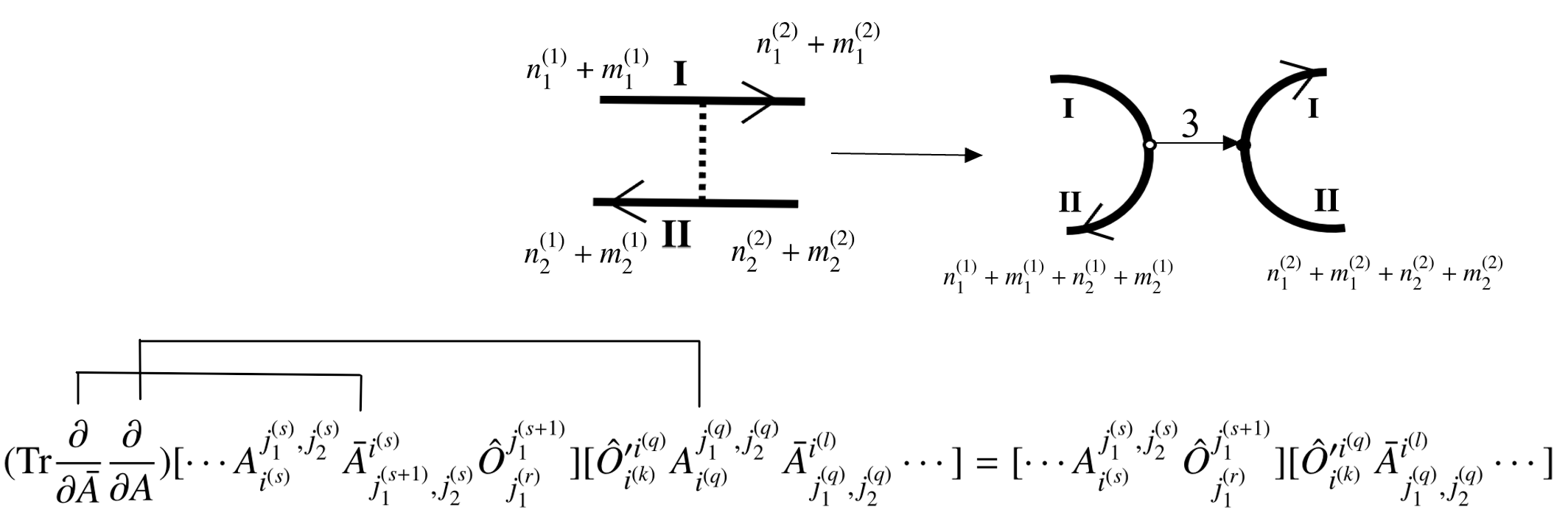}.
\end{eqnarray}

\subsection{Counting gauge-invariant operators}
Let us focus on the gauge invariants $\mathcal{T}_{\check{\sigma}}^{(n,m)}$ with fixed positive
integers~$n$,~$m$ in the graded ring $\mathcal{S}$.
$H_{n,m}$ acts on the left on $\mathcal{A}_{I}^{J_1,J_2}$ by simply swapping the tensors $A$ among
themselves and the tensors $B$ among themselves.  Similarly, $H_{n,m}$ acts on the right on
$\bar{\mathcal{A}}^{I}_{J_1,J_2}$ by swapping the tensors $\bar A$ among themselves and the tensors
$\bar B$ among themselves.
Let $h_1,\ h_2\in H_{n,m}$, thus
\begin{eqnarray}\label{permutation}
\mathcal{A}_{I}^{J_1,J_2}\cdot(\sigma_1,\sigma_2,\sigma_3)\cdot\bar{\mathcal{A}}^{I}_{J_1,J_2}
&=&
(\mathcal{A}_{I}^{J_1,J_2}\cdot h_1)\cdot(\sigma_1,\sigma_2,\sigma_3)\cdot(h_2 \cdot\bar{\mathcal{A}}^{I}_{J_1,J_2})
\nonumber\\&=&\mathcal{T}_{(h_1\sigma_1h_2, \ h_1\sigma_2h_2, \ h_1\sigma_3h_2)}^{(n,m)}.
\end{eqnarray}
We see that~$\mathcal{T}_{\check{\sigma}}^{(n,m)}$ and $\mathcal{T}_{(h_1\sigma_1h_2,\
h_1\sigma_2h_2,\  h_1\sigma_3h_2)}^{(n,m)}$ are indeed the same gauge-invariant operator.
 It implies that~$\sigma_i$, $i=1,2,3$, in~$\mathcal{T}_{\check{\sigma}}^{(n,m)}$
are characterized by the double coset~$H_{n,m}\setminus S_{n+m}\times S_{n+m}\times H_{n,m}/H_{n,m}$ \cite{jbgeloun1,jbgeloun2}.

For a permutation $p$ in $S_l$, the size of conjugacy class $|T_p|$ is given by
$|T_p|=\frac{l!}{Sym_l(p)}$, where $Sym_l(p)=\prod_{i=1}^N(i^{p_i})(p_i!)$ is the number of elements of
$S_l$ commuting with any permutation in the conjugacy class $T_p$, and $p_i$ gives the number of
cycles of size $i$ in $p$. Let us take~$h_2=gk$, where~$g\in S_n$,~$k\in S_m$. By Burnside's Lemma,
the number of elements in this double coset is
\begin{eqnarray}\label{nedc}
&&|H_{n,m}\setminus S_{n+m}\times S_{n+m}\times H_{n,m}/H_{n,m}|\nonumber\\
&=&\frac{1}{|H_{n,m}|^2}\sum_{h_1,h_2,\sigma_3\in H_{n,m}}\sum_{\sigma_1,\sigma_2\in S_{n+m}}\delta(h_1\sigma_1h_2\sigma_1^{-1})
\delta(h_1\sigma_2h_2\sigma_2^{-1})\delta(h_1\sigma_3h_2\sigma_3^{-1})\nonumber\\
&=&\frac{1}{|H_{n,m}|^2}\sum_{h_2\in H_{n,m}}\frac{n!}{Sym_n(g)}\cdot \frac{m!}{Sym_m(k)}\cdot
(Sym_{n+m}(h_2))^2\cdot Sym_n(g)\cdot Sym_m(k)\nonumber\\
&=&\frac{1}{|H_{n,m}|}\sum_{h\in H_{n,m}}(Sym_{n+m}(h))^2.
\end{eqnarray}
For the given~$h_2$ and~$h_1$ in $T_{h_2}$ in the second line of (\ref{nedc}),
$\delta(h_1\sigma_ih_2\sigma_i^{-1}),\ i=1, 2$
constraints $\sigma_i$ to be the permutations in $S_{n+m}$ commuting with
any permutation in $T_{h_2}$, which is exactly $Sym_{n+m}(h_2)$.
$\delta(h_1\sigma_3h_2\sigma_3^{-1})$ limits the sum over $\sigma_3$ to select only
the permutations which commute with any element of the conjugacy class of $h_2$ seen as an
element of $S_n \times S_m$. This yields
$Sym_{n}(g)Sym_{m}(k)$.

For example, when $n=m=1$, $H_{1,1}$ only contains the identity mapping $id=(1)(2)$, and~$Sym_{1+1}(id)=Sym_{2}(id)=1^2\cdot2!=2$, then
\begin{eqnarray}\label{case1}
|H_{1,1}\setminus S_2\times S_2\times H_{1,1}/H_{1,1}|=(Sym_{1+1}(id))^2=2^2=4.
\end{eqnarray}
For the case of $n=1$ and~$m=2$, $H_{1,2}=\{id=(1)(2)(3),(1)(23)\}$ in (\ref{nedc}), then~$Sym_{1+2}(id)=Sym_{3}(id)=1^3\cdot3!=6$,
~$Sym_{1+2}((1)(23))=Sym_{3}((1)(23))=1^1\cdot2^1\cdot1!\cdot1!=2$,
we have
\begin{eqnarray}\label{case2}
|H_{1,2}\setminus S_3\times S_3\times H_{1,2}/H_{1,2}|
=\frac{1}{2}((Sym_{1+2}(id))^2+(Sym_{1+2}((1)(23)))^2)
=\frac{1}{2}(6^2+2^2)=20.
\end{eqnarray}
Let us list more examples in Table $1$.
\begin{eqnarray*}
&&\includegraphics[width=7cm]{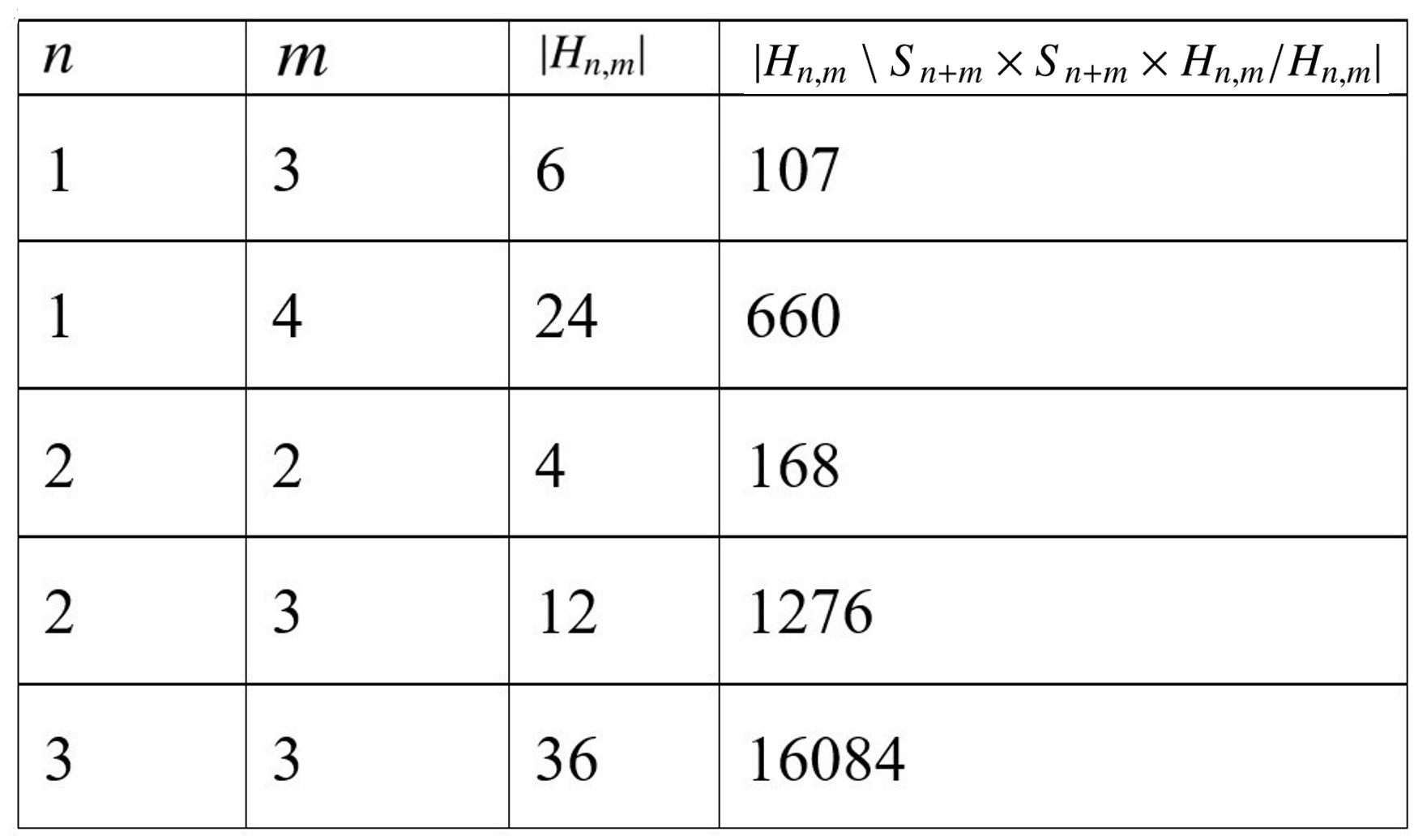}\nonumber\\
&&\text{Table 1. Examples with different $n$ and $m$ }.
\end{eqnarray*}

\begin{figure}[H]
\centering
\includegraphics[width=14cm]{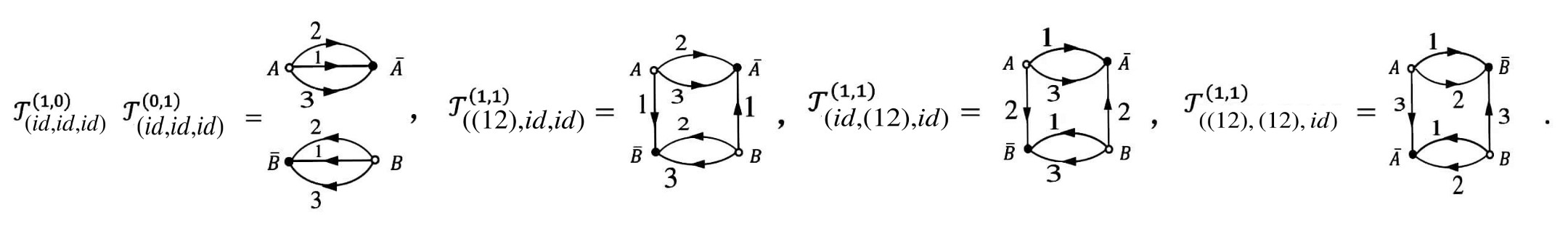}
\caption{Four operators in the case of $n=1$ and $m=1$ in (\ref{case1}).
White dots correspond to tensor fields~$A_{ {i}}^{{{j_1}},{ {j_2}}}$ and $B_{ {i}}^{{{j_1}},{
{j_2}}}$, black dots to~$\bar{A}^{ {i}}_{{{j_1}},{ {j_2}}}$ and~$\bar{B}^{ {i}}_{{{j_1}},{ {j_2}}}$;
lines of colors $1, 2$ and $3$ represent the contracting $i$, $j_1$ and $j_2$ indices, respectively.}
\label{2.7}
\end{figure}

\begin{figure}[H]
\centering
\includegraphics[width=14.5cm]{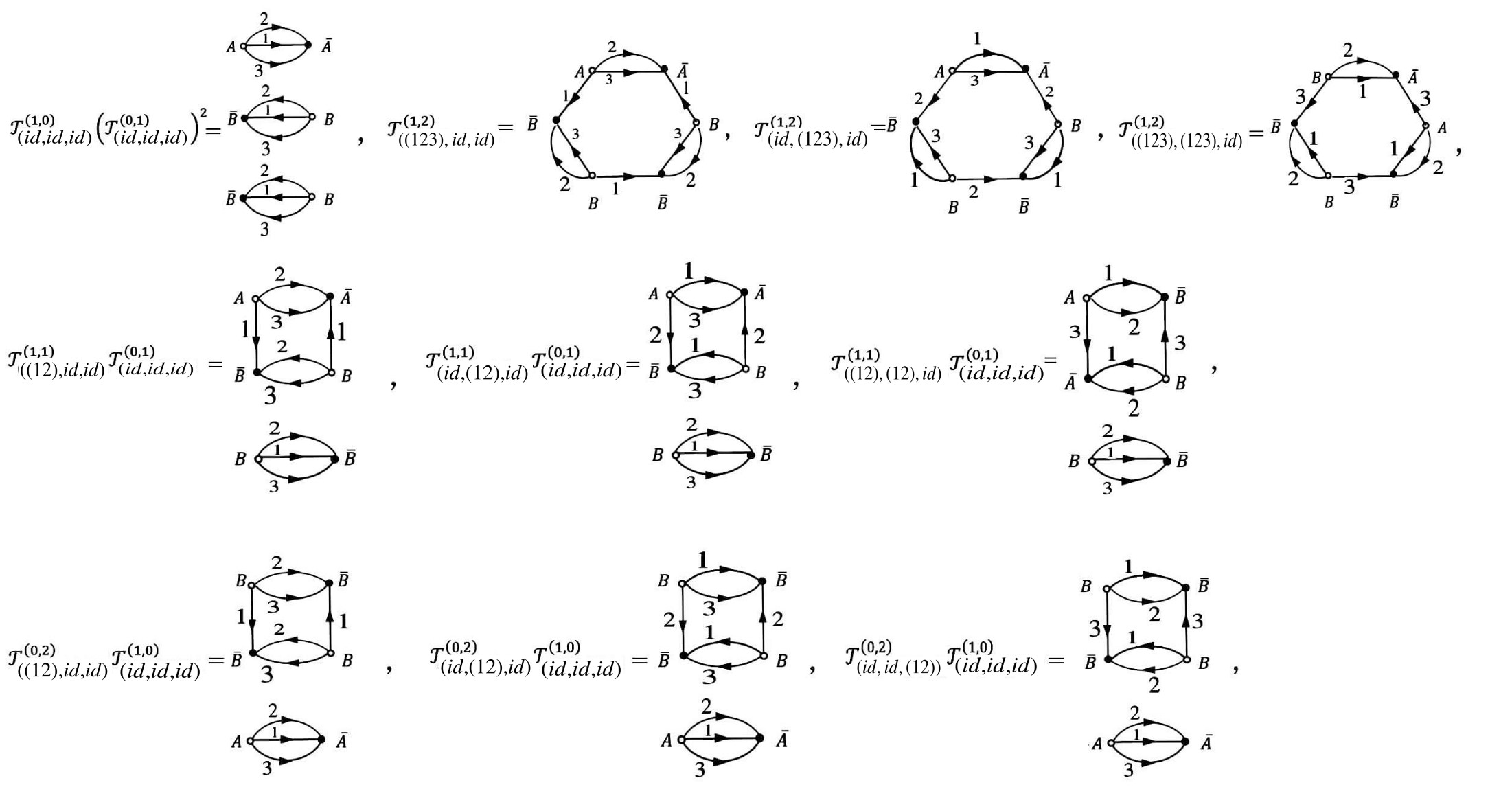}\\
\includegraphics[width=14.75cm]{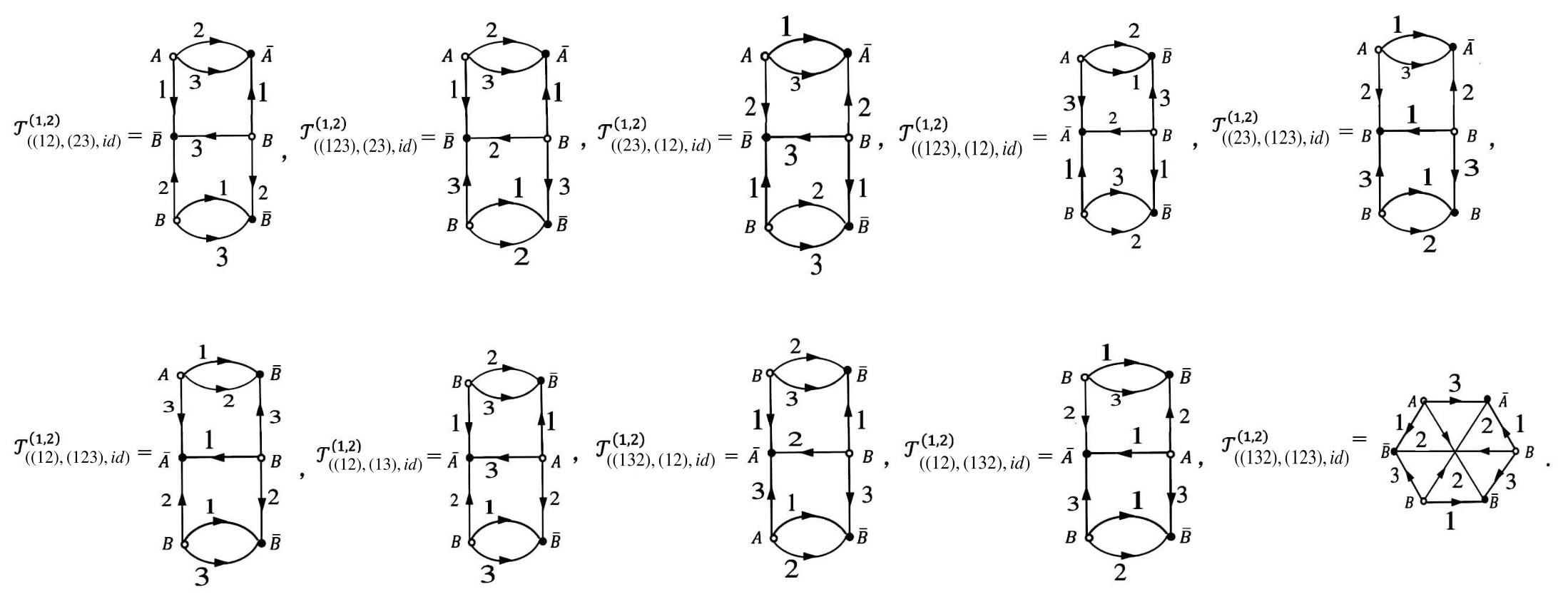}
\caption{Twenty operators in the case of $n=1$ and $m=2$ in (\ref{case2}).}
\label{2.8}
\end{figure}

We draw the operators of the cases (\ref{case1}) and (\ref{case2}) in Figs. \ref{2.7} and \ref{2.8}, respectively.

Thus the number of independent operators at each level-$l$ is
\begin{eqnarray}
\sharp_{l}=\sum_{n=1}^{l-1}\frac{1}{|H_{n,l-n}|}\sum_{h\in H_{n,l-n}}(Sym_{l}(h))^2+2\sum_{p\vdash l}Sym_{l}(p),
\end{eqnarray}
where the second sum below over~$p$ is performed over all partitions of~$l$ \cite{ItoyamaNPB2018}.
From the following relation between disconnected operators and connected operators,
\begin{eqnarray}\label{m+n-con-ope}
\eta(q)=1+\sum_{l=1}^{\infty}\sharp_{l}q^{l}=\textrm{PE}(\eta^{conn}(q))=
\textrm{PE}(\sum_{l=1}^{\infty}\sharp_{l}^{conn}q^{l})=\prod_{l=1}^{\infty}\frac{1}{(1-q^{l})^{\sharp_{l}^{conn}}},
\end{eqnarray}
 the number~$\sharp_{l}^{conn}$ can be read off from the plethystic logarithm
\begin{eqnarray}
\textrm{PLog}((\eta(q)))=\sum_{l=1}^{\infty}\sharp_{l}^{conn}q^l=\sum_{m=1}^{\infty}\frac{\mu(m)}{m}\log\eta(q^m),
\end{eqnarray}
where~$\mu(m)$ is the M\"{o}bius function
\begin{equation}
\mu(m)=\left\{
\begin{aligned}
&0,\quad &m \emph{ has at least one repeated prime factor,}\\
&1,\quad &m=1,\\
&(-1)^n, \quad &m\emph{ is a product of n distinct primes.}\\
\end{aligned}
\right.
\end{equation}

\subsection{Cut and join structure}

Let us now discuss the kernel and cokernel of the cut-and-join structure in the ring $\mathcal{S}$.
Due to the symmetry of $\Delta$ and $\tilde{\Delta}$ (or $\{ ,\}_A~ and ~\{ ,\}_B$),
we only focus on the cases $\Delta$ and $\{ ,\}_A$ here.

Since $\Delta$ maps all the operators at level-$(n+m)$ to those at level-$(n+m-1)$,
$\Delta$ inevitably has a kernel \cite{ItoyamaNPB2018}
\begin{eqnarray}
\Ker(\Delta)=\{\mathcal{T}_{\check{\sigma}}^{(n,m)}\in \mathcal{S}_{n+m}|\Delta(\mathcal{T}_{\check{\sigma}}^{(n,m)})=0\}\in \mathcal{S}_{n+m},
\end{eqnarray}
where $\mathcal{S}_{n+m}$ denotes the grading $n+m$ part of the ring $\mathcal{S}$.
Due to the cokernel, there are many operators which are not the descendants produced only by join operation of keystones.
It prevents the direct construction of the non-perturbed RG-complete partition function.

In what follows, we take the rings $\mathcal{S}_2$ and $\mathcal{S}_3$ as examples.
In the ring $\mathcal{S}_2$, there are three disconnected gauge-invariant operators
$(\mathcal{T}_{(id, id, id)}^{(1,0)})^2$,
$\mathcal{T}_{(id, id, id)}^{(1,0)}\mathcal{T}_{(id, id, id)}^{(0,1)}$,
$\mathcal{T}_{(id, id, id)}^{(0,1)}\mathcal{T}_{(id, id, id)}^{(0,1)}$
and nine connected gauge-invariant operators
$\mathcal{T}_{((12), id, id)}^{(2,0)}$,
$\mathcal{T}_{((12), id, id)}^{(0,2)}$,
$\mathcal{T}_{(id, (12), id)}^{(2,0)}$,
$\mathcal{T}_{(id, id, (12))}^{(2,0)}$,
$\mathcal{T}_{(id, id, (12))}^{(2,0)}$,
$\mathcal{T}_{(id, id, (12))}^{(0,2)}$,
$\mathcal{T}_{((12), id, id)}^{(1,1)}$,
$\mathcal{T}_{(id, (12), id)}^{(1,1)}$,
$\mathcal{T}_{((12), (12), id)}^{(1,1)}$.
The cut operation $\Delta$ takes operators from $\mathcal{S}_2$  to level-$1$ operators
$\mathcal{T}_{(id, id , id)}^{(1,0)} ~and~\mathcal{T}_{(id , id, id)}^{(0,1)}$.
In this case, through calculations we know that $\Delta$ has a kernel (of codimension two) with dimension ten
\begin{eqnarray}
\ker^{(2)}(\Delta)&=&\span\{
\mathcal{T}_{((12), id , id)}^{(0,2)},
\mathcal{T}_{(id , (12), id)}^{(0,2)},
\mathcal{T}_{(id, id , (12))}^{(0,2)},
\mathcal{T}_{(id, id , id)}^{(0,1)}\mathcal{T}_{(id, id , id)}^{(0,1)},
\nonumber\\
&&
(\beta+1)\mathcal{T}_{((12),  id, id)}^{(2,0)}
- {\alpha}(\mathcal{T}_{(id, id , id)}^{(1,0)})^2,
~
(\beta+1)\mathcal{T}_{(id,  (12), id)}^{(2,0)}
-\tilde{\alpha}(\mathcal{T}_{(id, id , id)}^{(1,0)})^2,
\nonumber\\
&&
(\beta+1)\mathcal{T}_{(id,  id, (12))}^{(2,0)}
-\hat{\alpha}(\mathcal{T}_{(id, id , id)}^{(1,0)})^2,
~
{N_2} {N_3}\mathcal{T}_{((12),  (12), id)}^{(1,1)}
- {N_3}\mathcal{T}_{((12),  id, id)}^{(1,1)},
\nonumber\\
&&
 {N_1} {N_3}\mathcal{T}_{((12),  (12), id)}^{(1,1)}
- {N_3}\mathcal{T}_{(id,  (12), id)}^{(1,1)},
~
\beta\mathcal{T}_{((12),  (12), id)}^{(1,1)}
- {N_3}\mathcal{T}_{(id, id , id)}^{(1,0)}\mathcal{T}_{(id, id , id)}^{(0,1)}
\},
\end{eqnarray}
where $\alpha=N_1+N_2N_3$, $\beta=N_1N_2N_3$,
 $\tilde{\alpha}=N_2+N_1N_3$ and $\hat{\alpha}=N_3+N_1N_2$.

In the following discussion, we omit the operators involving only field $B$ and $\bar{B}$,
since they are obviously included in kernel.
Thus in $\mathcal{S}_3$, we only consider the gauge-invariant operators
$\mathcal{T}_{((123), (132) , id)}^{(3,0)}$, $\mathcal{T}_{((123), (132) , id)}^{(2,1)}$ and the level-$3$
operators produced by join operation $\{ ,\}_A$ (see Table 2).

\begin{eqnarray*}
  &&\includegraphics[width=15cm,height=6cm]{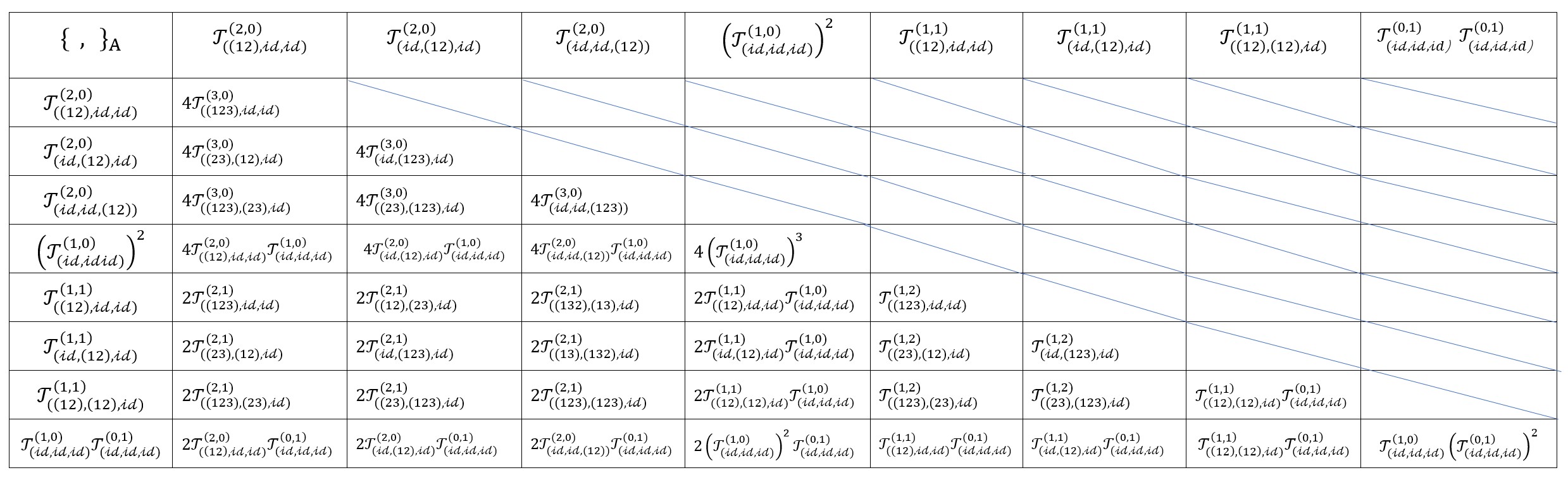}\nonumber\\
&&\text{Table 2. Level-$3$ operators produced by join operation $\{~,\}_A$}.
 \end{eqnarray*}
The cut operation $\Delta$ acting on the above level-$3$ gauge-invariant operators are listed in (\ref{cutlevel}).
Then we can give the kernel $\ker^{(3)}(\Delta)$ with dimension $29$ (see (\ref{cutkernel})) and cokernel as follows
\begin{eqnarray}
\Coker^{(3)}(\{,\}_A)=\span\{\mathcal{T}_{((123), (132), id)}^{(3,0)},
\mathcal{T}_{((123), (132), id)}^{(2,1)}\}.
\end{eqnarray}

\subsection{Join operators and Fredkin spin chain}

The Fredkin spin chain \cite{DellAnna, Salberger} is a spin chain of length-$2n$, where up and down spin degrees of
freedom with multiplicity (called as color) $s$ are assigned at each of the lattice sites $\{1,2,\cdots,2n\}$.
There is the connection between the Fredkin spin chain and large $N$ matrix models \cite{Fumihiko}.
The Hamiltonian of Fredkin spin chain is
\begin{eqnarray}\label{hamilton}
H_{F,s}&=&\frac{1}{2}\sum_{j=1}^{2n-2}\sum_{k_1,k_2,k_3=1}^s
\{(|u_j^{k_1},u_{j+1}^{k_2},d_{j+2}^{k_3}\rangle-|u_j^{k_1},d_{j+1}^{k_2},u_{j+2}^{k_3}\rangle)
(\langle u_j^{k_1},u_{j+1}^{k_2},d_{j+2}^{k_3}|-\langle u_j^{k_1},d_{j+1}^{k_2},u_{j+2}^{k_3}|)\nonumber\\
&&+(|u_j^{k_1},d_{j+1}^{k_2},d_{j+2}^{k_3}\rangle-|d_j^{k_1},u_{j+1}^{k_2},d_{j+2}^{k_3}\rangle)
(\langle u_j^{k_1},d_{j+1}^{k_2},d_{j+2}^{k_3}|-\langle d_j^{k_1},u_{j+1}^{k_2},d_{j+2}^{k_3}|)\}\nonumber\\
&&+\sum_{j=1}^{2n-1}\sum_{k\neq l}\{|u_j^{k},d_{j+1}^{l}\rangle\langle u_j^{k},d_{j+1}^{l}|
+\frac{1}{2}(|u_j^{k},d_{j+1}^{k}\rangle-|u_j^{l},d_{j+1}^{l}\rangle)(\langle u_j^{k},d_{j+1}^{k}|
-\langle u_j^{l},d_{j+1}^{l}|)\}\nonumber\\
&&+\sum_{k=1}^{s}\{|d_{1}^{k}\rangle\langle d_{1}^{k}|+|u_{2n}^{k}\rangle\langle u_{2n}^{k}|\},
\end{eqnarray}
where we express the up- and down-spin states with color~$k\in\{1,2,\cdots,s\}$ at the site~$i $ as~$|u_{i}^k\rangle$
and~$|d_{i}^k\rangle$, respectively.
For the von Neumann entropy of the Fredkin model, it shows a
 square-root violation of the area law, and the scaling behavior of the first gap \cite{DellAnna}.

In this paper, we only focus on the case of $s=2$ in (\ref{hamilton}). Thus the up- and down-spin
states can be represented as arrows with color degrees of freedom in the
two-dimensional~$(x,y)$-plane pointing to~$(1,1)$ (up-step) and~$(-1,-1)$ (down-step), respectively.
For the Hamiltonian (\ref{hamilton}), it has a unique ground state at zero energy, which is
superposition of spin configurations with equal weight. Each spin configuration appearing
in the superposition is identified with each path of length-$2n$ Dyck walks. For the Dyck walks,
they are random walks starting at the origin, ending at~$(2n, 0)$ and restricted to the region $y
\geqslant0$. Furthermore the color of each up-step should be matched with that of the down-step
subsequently appearing at the same height.

The ground state of the Fredkin spin chain is given by
\begin{eqnarray}
|P_{F,2n,2}\rangle=\frac{1}{\sqrt{N_{F,2n,2}}}\sum_{\omega\in P_{F,2n,2}}|\omega\rangle,
\end{eqnarray}
where~$ P_{F,2n,2}$ denotes the formal sum of length-$2n$ colored Dyck walks,~$\omega$ runs over
monomials appearing in~$ P_{F,2n,2}$, and~$N_{F,2n,2}$ counts the number of the length-$2n$
colored Dyck walks
\begin{eqnarray}\label{Catalan}
N_{F,2n,2}=2^nN_{F,2n}
=\frac{2^{n}}{n+1}\binom{2n}{n},
\end{eqnarray}
where~$N_{F,2n}$ denotes the~$n$-th Catalan number.

\begin{figure}[H]
\centering
\includegraphics[width=11cm]{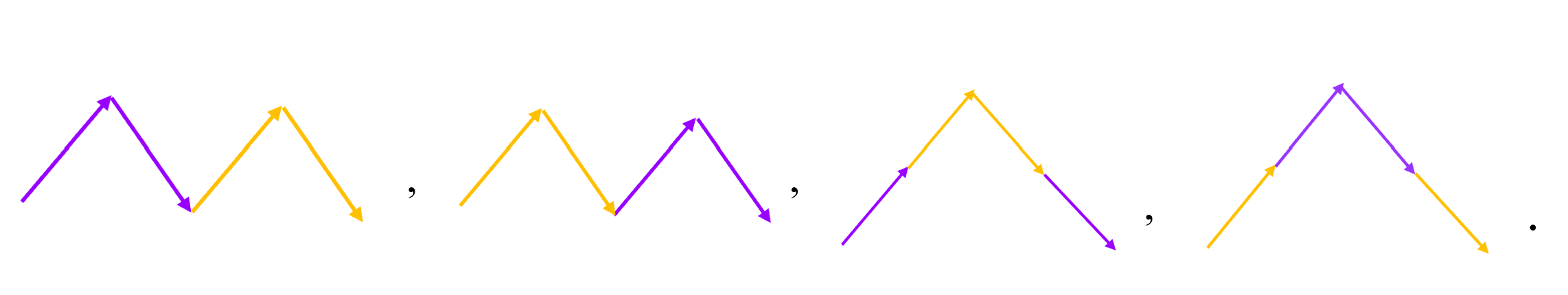}
\caption{ Colored Dyck walks in the summand of (\ref{225}). }
\label{colordyck}
\end{figure}

Let us take~$n=2$ as an example,
we have~$N_{F,4,2}=8$ and
\begin{eqnarray}\label{225}
|P_{F,2n,2}\rangle=\frac{1}{2\sqrt{2}}(|u^1d^1u^2d^2\rangle+|u^2d^2u^1d^1\rangle+|u^1u^2d^2d^1\rangle+|u^2u^1d^1d^2\rangle).
\end{eqnarray}
The four states of the summand can be drawn as the colored Dyck walks (see Fig. \ref{colordyck}).

In the previous subsection, we have presented the tree operators which are constructed by the keystone operators
through the join operation.
Since the gauge invariants $\mathcal{T}_{\check{\sigma}}^{(n,m)}$ with fixed positive integers $n,
m$ constitute the double coset $H_{n,m}\setminus S_{n+m}\times S_{n+m}\times H_{n,m}/H_{n,m}$, there
is $\check{\sigma}\cong( h_1\sigma_1h_2,h_1\sigma_2 h_2, h_1\sigma_3 h_2),$
where~$h_1,h_2\in H_{n,m}$. If there exist~$h_1,h_2$~such that $h_1\sigma_1h_2 = id$
and~$h_1\sigma_2 h_2$ is in the conjugacy class of $(1,2,\cdots,n+m)$, we obtain the conjugacy class
in which the Feynman diagrams are \textbf{I}-\textbf{II} cycles with shortcuts of color $3$.
When the shortcuts are disjoint, the Feynman diagrams represent the tree operators (see Figs.
\ref{alltree}(c),(d)).

\begin{figure}[H]
\centering
\includegraphics[width=13cm,height=5.25cm]{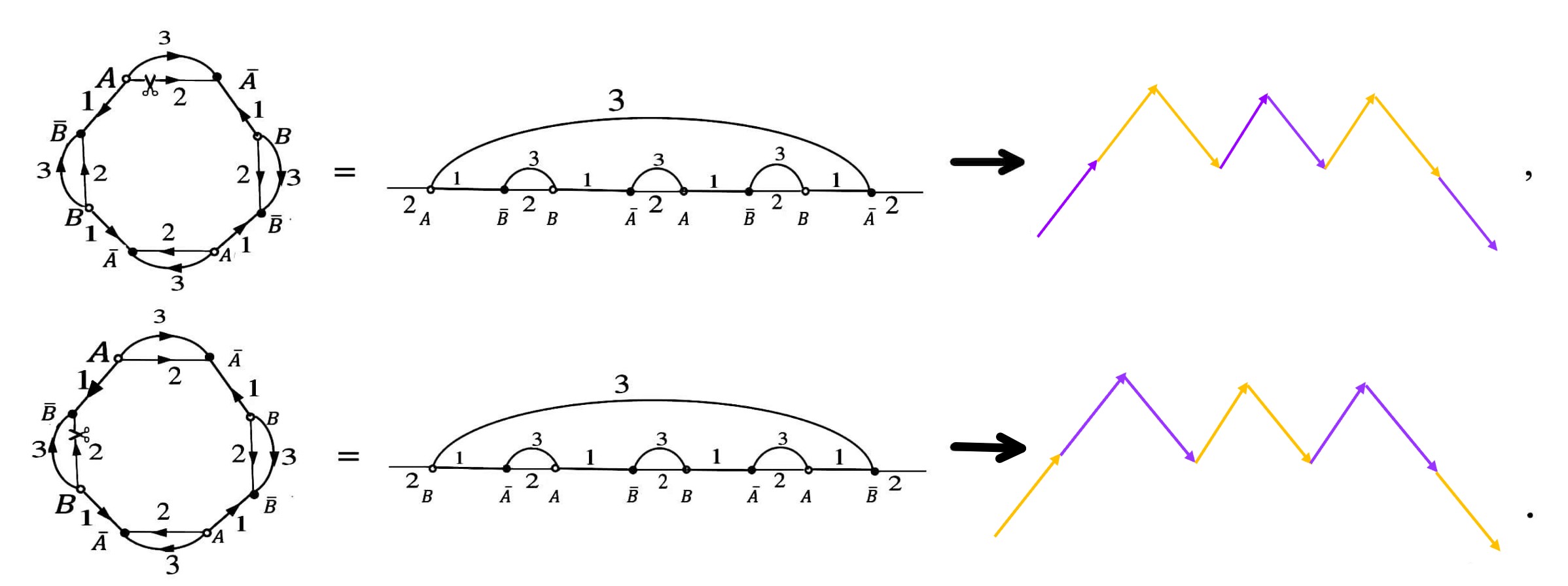}
\caption{ Correspondence between $\mathcal{T}_{((1324), id , id)}^{(2,2)}$ and length-8 Dyck walks.}
\label{rotation}
\end{figure}

Let us now establish the correspondence between two colored Dyck walks with length-$2(n+m)$ in
Fredkin spin chain and tree operators. By cutting tree operators from lines of color $2$, it gives a
\textbf{I} or \textbf{II} segment with elements connected by disjoint  lines of color $3$, where
each  line of color $3$ may correspond a up- and down-step. If the  line of color $3$ connects A
and~$\bar{A}$, we paint the step with purple, otherwise with yellow.  We make the correspondence rule as follows:

(i) For the tree operators with rotational symmetry, we cut the fundamental domain of Feynman diagrams.
For example, the Feynman diagram of~$\mathcal{T}_{((1324), id , id)}^{(2,2)}$ has 2-fold rotation
symmetry. Thus we only need to cut two lines color $2$ connecting $A$ and~$\bar{A}$, B
and~$\bar{B}$, respectively (see left-hand side of Fig. \ref{rotation}). The segments \textbf{I} and
\textbf{II} are shown in the middle parts of Fig. \ref{rotation}. Each line of color $3$ in chains corresponds
a up- and down-step in Dyck walks (see right-hand side of Fig. \ref{rotation}).

\begin{figure}[H]
\centering
\includegraphics[height=1.25cm]{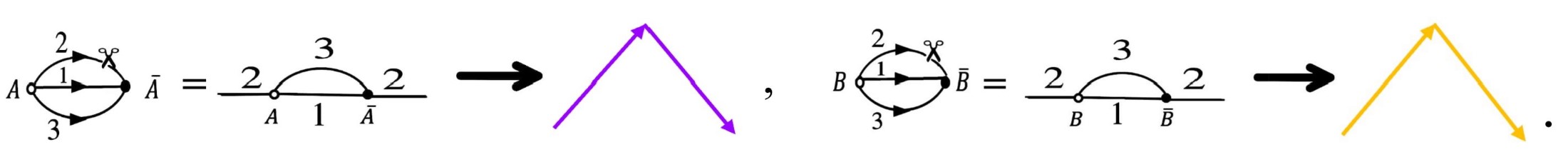}
\caption{ Correspondence between $\mathcal{T}_{(id, id , id)}^{(1,0)}$,
$\mathcal{T}_{(id, id , id)}^{(0,1)}$ and length-2 Dyck walks.}
\label{trivial}
\end{figure}

\begin{figure}[H]
\centering
\includegraphics[width=8cm]{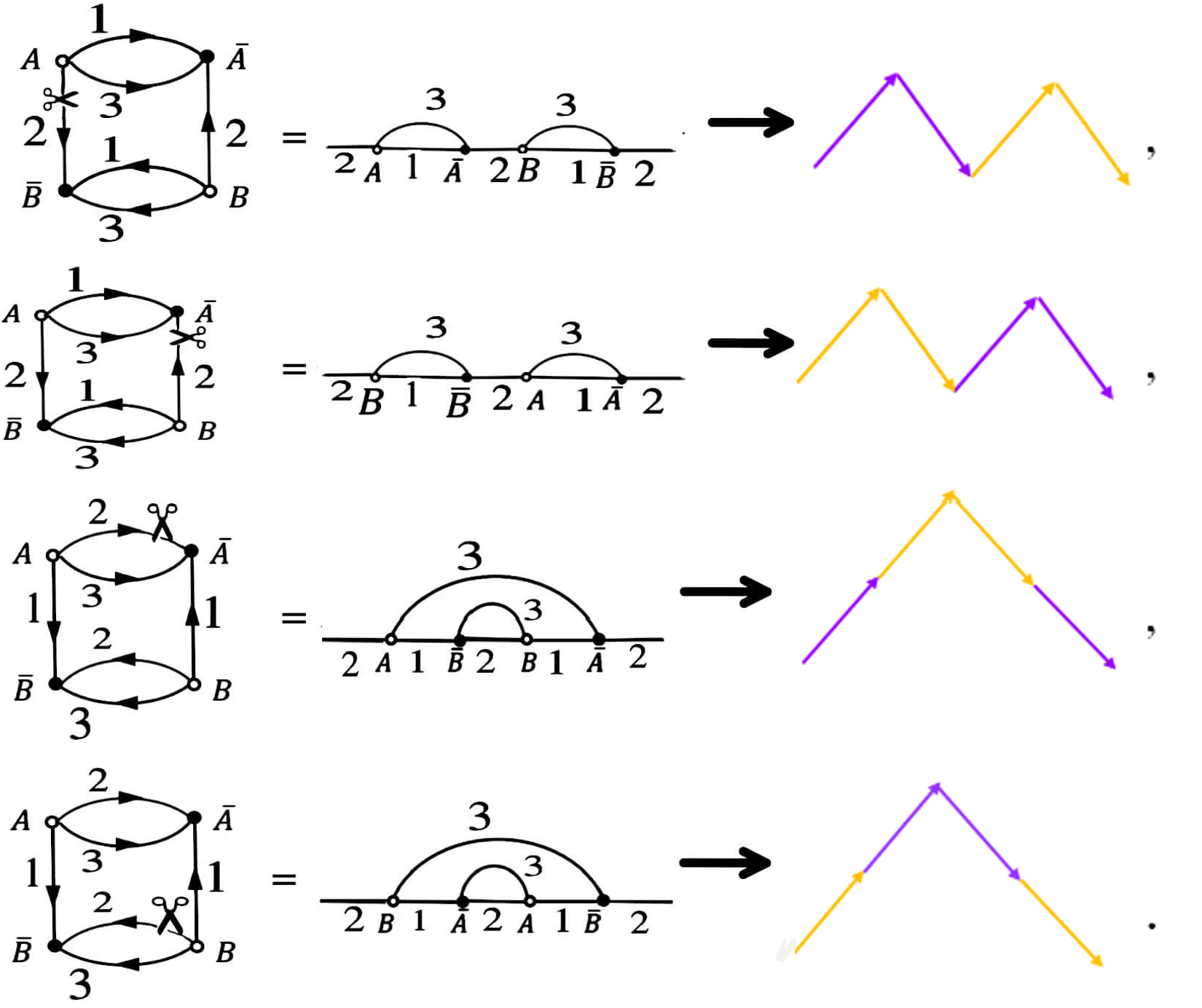}
\caption{ Correspondence between $\mathcal{T}_{(id, (12) , id)}^{(1,1)}$, $\mathcal{T}_{((12), id ,
id)}^{(1,1)}$ and length-4 Dyck walks.}
\label{norotation}
\end{figure}

(ii) For the tree operators without rotational symmetry, we cut them from each line of color $1$ or $2$
(see examples in Figs. \ref{trivial} and \ref{norotation}).

Here for the rotational symmetry of tree operators $\mathcal{T}_{\check{\sigma}}^{(n,m)}$, it means
that there exist the disjoint $d$-cycles $\omega_1,\cdots,\omega_j$, $(d\times j=n+m)$,  such that
the following relations hold
\begin{eqnarray}
\mathcal{T}_{(\omega_1\omega_2\cdots\omega_j\circledast\sigma_1\circledast(\omega_1\omega_2\cdots\omega_j)^{-1}
, \omega_1\omega_2\cdots\omega_j\circledast \sigma_2 \circledast(\omega_1\omega_2\cdots\omega_j)^{-1}
, \omega_1\omega_2\cdots\omega_j\circledast \sigma_3\circledast(\omega_1\omega_2\cdots\omega_j)^{-1})}^{(n,m)}
=\mathcal{T}_{\check{\sigma}}^{(n,m)},
\end{eqnarray}
where the symbol ``$\circledast$" represents the composition of maps.
Then we call the maximum value of $d$ the fold of rotational symmetry
of $\mathcal{T}_{\check{\sigma}}^{(n,m)}$.

\begin{figure}[H]
\centering
\includegraphics[width=8cm,height=4cm]{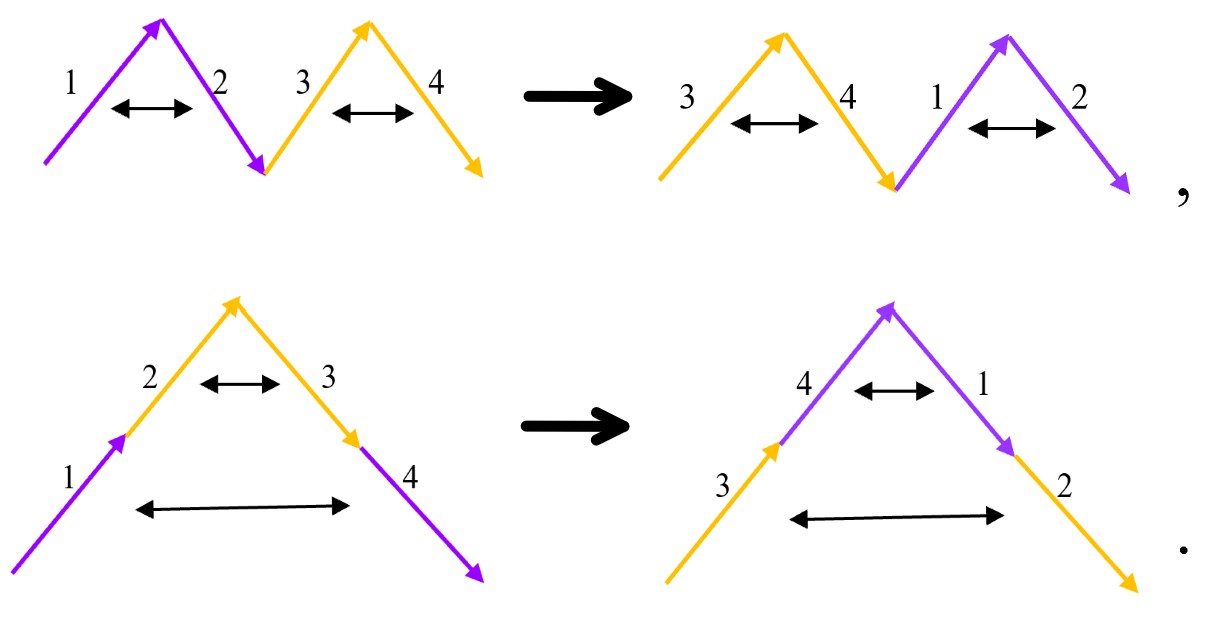}
\caption{ Two examples of moving the first two steps of Dyck walks to the back in turn.}
\label{Dyckwalks}
\end{figure}

From such correspondence, we see that $N_{F,2(n+m),2}$ counts the weight~$\frac{n+m}{r}$ of tree operators
with level-$(m+n)$, where~$r$ denotes the fold of rotational symmetry,
\begin{eqnarray}\label{NF}
N_{F,2(n+m),2}&=&2^{(n+m)}N_{F,2(n+m)}=\frac{2^{n+m}}{n+m+1}\binom{2(n+m)}{n+m}
\nonumber\\
&=&\sum_{i=1}^{\sharp_{n+m}^{tree}}\left(\text{weight of
 tree operator~$\mathcal{T}^{(n,m)}_{\check{\alpha_i}}$}\right).
\end{eqnarray}
Then from $(\ref{NF})$, we obtain a partition $(n_1,n_2,\cdots,n_{\sharp_{n+m}^{tree}})$ of $N_{F,2(n+m),2}$,
where $\sharp_{n+m}^{tree}$ denotes the number of tree operators with level-$(n+m=l)$.
This partition gives a classification of length-$2(n+m)$ Dyck walks. Let us number the steps of Dyck walks.
Then we move the first two steps to the back in turn and remain the matching relation (see examples in
Fig. \ref{Dyckwalks}). By repeating the process above, we may obtain all Dyck walks belonging the same class,
i.e., corresponding to the same tree operator cutted from different lines of color $2$.

Based on the classification Dyck walks, it is easy to give the number of tree operators with level-$l$
\begin{eqnarray}
\sharp_{l}^{tree}=
\frac{N_{F,2l,2}+\sum_{i=1}^{l-1}\sum_{r}\frac{r-1}{r}ll_{r}+2[2(l-1)+\sum_{q}\frac{{q}-1}{q}ll_{q}]}{l},
\end{eqnarray}
where $r$ is the common factor of $i$ and $l-i$, $q$ is the factor of $l$, $l_r$ and $l_q$ count the number
of Dyck walks with length $r$ and $q$, respectively.

Since the number of connected operators with level-$l$, $\sharp_{l}^{conn}$, can be calculated from
$(\ref{m+n-con-ope})$, we have the number of connected loop operators with level-$l$,
$\sharp_{l}^{loop}=\sharp_{l}^{conn}-\sharp_{l}^{tree}$.

\begin{figure}[H]
\centering
\includegraphics[width=7cm,height=4cm]{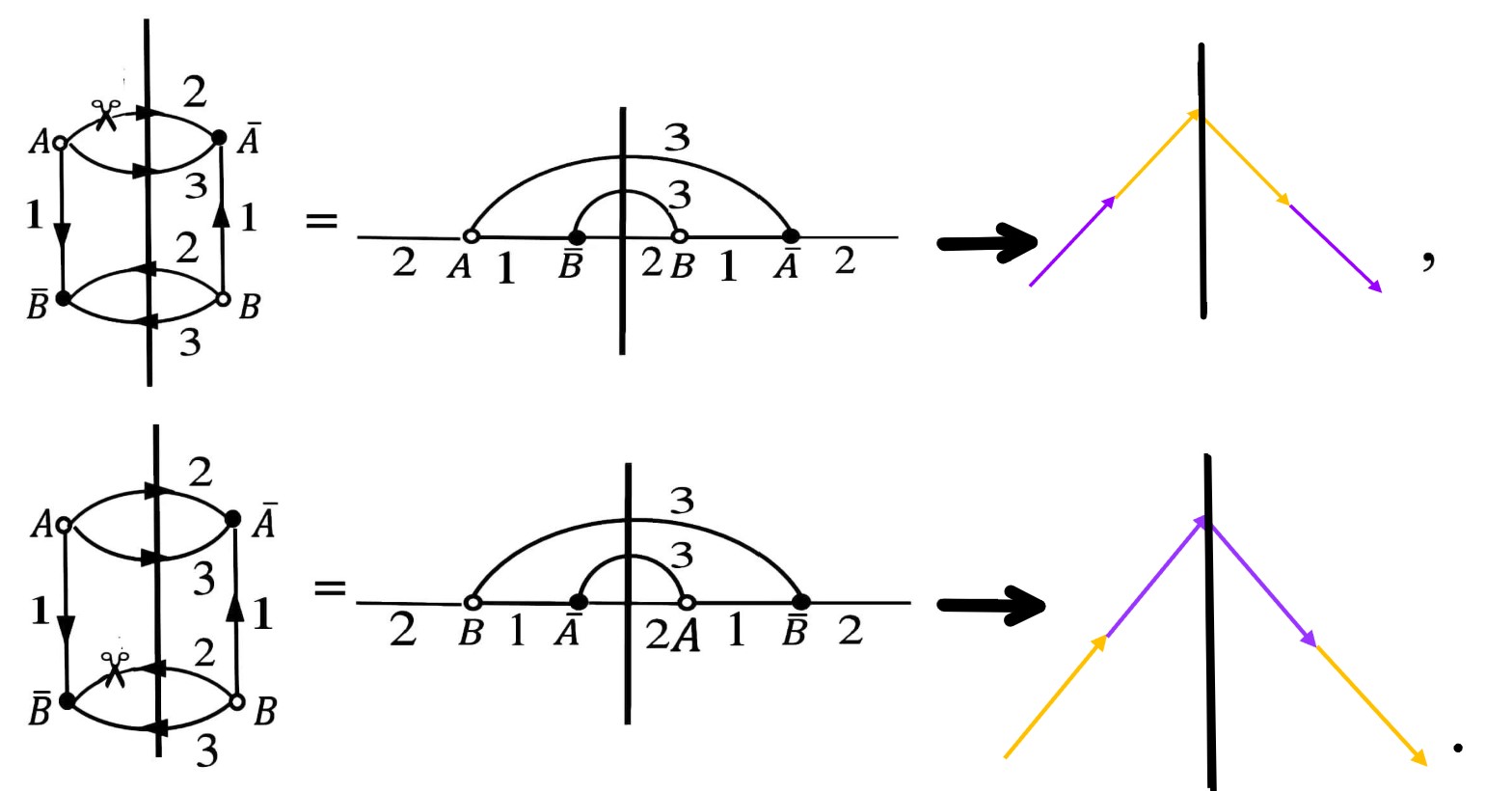}
\caption{Divide the cutted Feynman diagram of $\mathcal{T}_{((12), id , id)}^{(1,1)}$
into two parts, and two colored Dyck walks of two subsystems in the case of~$n=m=1$, $r=0$ and $h=2$.}
\label{cutteddiag}
\end{figure}

To divide the cutted Feynman diagram into two parts, we draw a vertical line passing the cutted line.
It gives $(n+m-r)$ and $(n+m+r)$ fields in the left and right parts, respectively.
Thus the corresponding two colored Dyck walks are divided into two subsystems,~$\mathcal{A}$ with $(n+m-r)$, and~$\mathcal{B}$
with $(n+m+r)$ spins, respectively. The spin configurations in subsystem $\mathcal{A}$ correspond to a part of colored Dyck paths
from the origin to $(n+m + r, h)$ in the~$(x, y)$-plane, and the spin configurations in subsystem $\mathcal{B}$ correspond to the
paths from $(n+m + r, h)$ to $(2(n+m), 0)$. The part of $\mathcal{A}$ has $h$ unmatched up-steps that are supposed to be matched
across the boundary with $h$ unmatched down-steps in the part of $\mathcal{B}$.
For example, the cutted Feynman diagrams of $\mathcal{T}_{((12), id , id)}^{(1,1)}$ correspond to the
Dyck walks of~$n=m=1$, $r=0$ and $h=2$ (see Fig. \ref{cutteddiag} ).

Let us denote the number of the paths in subsystem $\mathcal{A}$ as~$N_{F,n+m+r,2}^{(0\rightarrow h)}$,
and the number of the paths in subsystem $\mathcal{B}$ as $N_{F,n+m-r,2}^{(h\rightarrow 0)}$.
The correspondence implies that, $N_{F,n+m+r,2}^{(0\rightarrow h)}N_{F,n+m-r,2}^{(h\rightarrow 0)}$
counts the number of cutted Feynman diagrams with~$h$ intersection points of the vertical line and  lines of color $3$,
where
\begin{eqnarray}
N_{F,n+m+r,2}^{(0\rightarrow h)}=N_{F,n+m-r,2}^{(h\rightarrow 0)}=2^{\rho}N_{F,n+m+r}^{(h)}
=\left\{\begin{array}{cl}
   2^{\rho}\frac{h+1}{\rho+1}\binom{n+m+r}{\rho}, & \rho\in \mathbb{N},\\
  0,   & otherwise.
 \end{array}\right.
\end{eqnarray}

The entanglement entropy of the quantum system divided into two systems is given by
\begin{eqnarray}
s_{F,A}=-\sum_{h=0}^{n+m-|r|}2^hp_{F,n+m+r,n+m-r,2}^{(h)}\ln p_{F,n+m+r,n+m-r,2}^{(h)},
\end{eqnarray}
where
\begin{eqnarray}\label{ph}
p_{F,n+m+r,n+m-r,2}^{(h)}=2^{-2h}\frac{N_{F,n+m+r,2}^{(0\rightarrow h)}
N_{F,n+m-r,2}^{(h\rightarrow 0)}}{N_{F,2(n+m),2}}.
\end{eqnarray}
For the entanglement entropy of the Fredkin spin chain, it exhibits a nonlogarithmic violation of the area law,
which is beyond logarithmic scaling in the ordinary critical systems \cite{DellAnna}.

\begin{figure}[H]
\centering
\includegraphics[width=12.5cm,height=3cm]{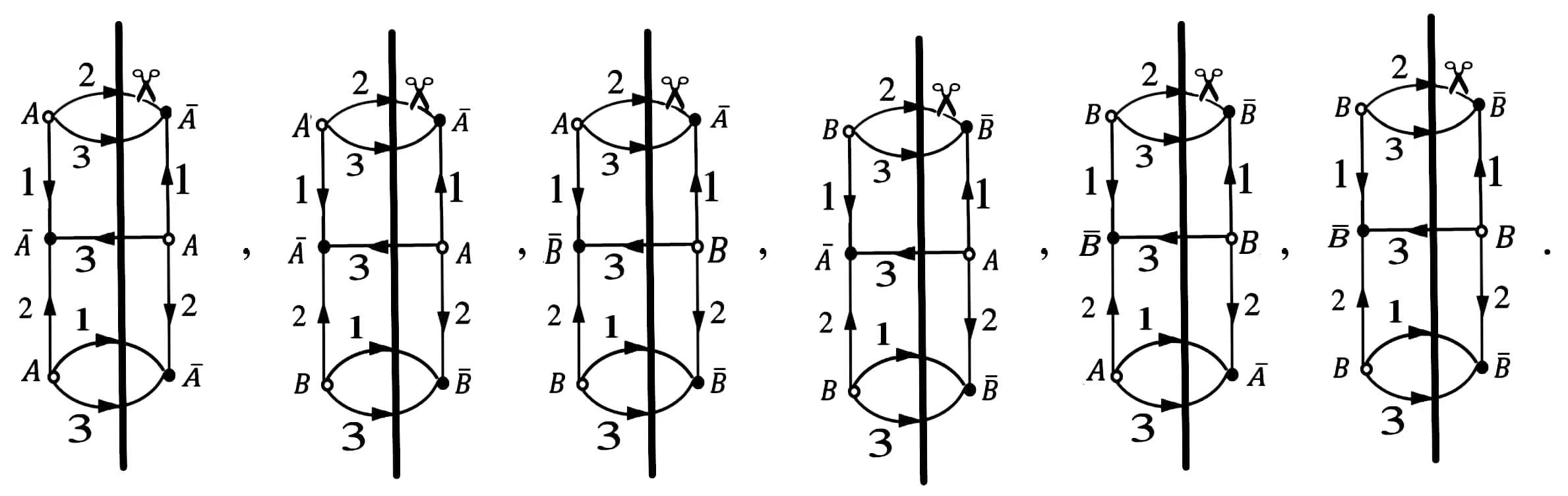}
\caption{The cutted Feynman diagrams contributing to~$p_{F,3,3,2}^{(3)}$. }
\label{SFA}
\end{figure}

The ratio (\ref{ph}) plays an important role in the entanglement entropy.
Note that~$p_{F,n+m+r,m+n-r,2}^{(n+m-|r|)}$ is always the smallest in all~
$p_{F,n+m+r,n+m-r,2}^{(h)}$ for~$n+m\geqslant 2$,
since there is only the contribution from the diagrams
with vertical line intersecting all possible lines of color $3$ (see examples in Fig. (\ref{SFA})).
It prohibits all possible~$p_{F,n+m+r,n+m-r,2}^{(h)}$ to be equal.
Hence we conclude that the following relation does not hold
\begin{eqnarray}\label{}
s_{F,A}=-\sum_{h=0}^{n+m-|r|}2^hp_{F,n+m+r,n+m-r,2}^{(h)}\ln p_{F,n+m+r,n+m-r,2}^{(h)}
=-\sum_{i=1}^{k}2^h(2^{-2h}\frac{1}{k})\ln(2^{-2h}\frac{1}{k})=2^{-h}\ln 2^{2h}k,
\end{eqnarray}
where~$k=[n+m-|r|]+1$ due to the construction of tree operator.
Thus we explain the entanglement scaling beyond logarithmic scaling in the ordinary critical systems
from the viewpoint of tensor model here.

\section{A two-tensor model with order-3}
\subsection{$W$-representation of the two-tensor model with order-3}

We may choose keystone operators constructed from $(\ref{keystoneope})$ to generate
a renormalization group completed two-tensor model with order-3
\begin{eqnarray}\label{rainbowrg}
Z_{AB}&=&
\int dA d\bar A \int dB d\bar B\exp(-\Tr A\bar A-\Tr B\bar B
+\sum_{n+m=1}^{\infty}\sum_{\check{\sigma} |\  \level(\check{\sigma})=n+m}N^{-2(n+m)}t_{\check{\sigma}}^{(n,m)}
\mathcal{T}_{\check{\sigma}}^{(n,m)}),
\nonumber\\
&=&
\int dA d\bar A \int dB d\bar B\exp(- \Tr A\bar A-\Tr B\bar B
+N^{-2}t_{(id, id , id)}^{(1,0)}\mathcal{T}_{(id, id , id)}^{(1,0)}
+N^{-2}t_{(id, id , id)}^{(0,1)}\mathcal{T}_{(id, id , id)}^{(0,1)}
\nonumber\\
&&
+\sum_{k=2}^{\infty} N^{-2k}t_{((12\cdots k), id , id)}^{(k,0)}
      \mathcal{T}_{((12\cdots k), id , id)}^{(k,0)}
+\sum_{k=2}^{\infty} N^{-2k}t_{((12\cdots k), id , id)}^{(0,k)}
      \mathcal{T}_{((12\cdots k), id , id)}^{(0,k)}
\nonumber\\
&&
+\sum_{k=2}^{\infty} N^{-2k}t_{(id, (12\cdots k) , id)}^{(k,0)}
      \mathcal{T}_{(id, (12\cdots k) , id)}^{(k,0)}
+\sum_{k=2}^{\infty} N^{-2k}t_{(id, (12\cdots k) , id)}^{(0,k)}
      \mathcal{T}_{(id, (12\cdots k) , id)}^{(0,k)}
\nonumber\\
&&
+\sum_{k=2}^\infty N^{-2k}t_{(id, (12\cdots k) , id)}^{(k,0)}
      \mathcal{T}_{(id, (12\cdots k) , id)}^{(k,0)}
+\sum_{k=2}^\infty N^{-2k}t_{(id, (12\cdots k) , id)}^{(0,k)}
      \mathcal{T}_{(id, (12\cdots k) , id)}^{(0,k)}
+\cdots),
\end{eqnarray}
where the measure is induced by the norm~$\parallel \delta A\delta B \parallel^2
=\delta A_{ {i}}^{j_1, { j_{2}}}
\delta \bar{A}^{ {i}}_{ {j_1}, { j_{2}}}
\delta B_{ {i}}^{ {j_1}, { j_{2}}}
\delta \bar{B}^{ {i}}_{ {j_1}, { j_{2}}}$.

By requiring that the partition function (\ref{rainbowrg}) is invariant under the transformation
$A\rightarrow A+\delta A$,
where $\delta A=\dsum_{a+b=1}^{\infty}\dsum_{\check{\alpha}|\ \level(\check{\alpha})=a+b}N^{-2(a+b)}t_{\check{\alpha}}^{(a+b)}\dfrac{\partial
\mathcal{T}_{\check{\alpha}}^{(a,b)}}{\partial \bar A}$, we have
\begin{eqnarray}\label{wardA}
&&\Big[-\dsum_{a+b=1}^{\infty}\dsum_{\check{\alpha}|\ \level(\check{\alpha})=a+b}aN^{-2(a+b)}t_{\check{\alpha}}^{(a,b)} \mathcal{T}_{\check{\alpha}}^{(a,b)}
+\dsum_{a+b=1}^{\infty}\dsum_{\check{\alpha}|\ \level(\check{\alpha})=a+b}N^{-2(a+b)}t_{\check{\alpha}}^{(a,b)}\Delta \mathcal{T}_{\check{\alpha}}^{(a,b)}
\nonumber\\
&&+\dsum_{\substack{a+b=1,\\n+m=1}}^{\infty}\dsum_{\substack{\check{\alpha}|\ \level(\check{\alpha})=a+b,
\\ \check{\sigma}|\ \level(\check{\sigma})=n+m}}
N^{-2(a+b)}N^{-2(n+m)}t_{\check{\alpha}}^{(a,b)}t_{\check{\sigma}}^{(n,m)}
\{\mathcal{T}_{\check{\sigma}}^{(n,m)},\mathcal{T}_{\check{\alpha}}^{(a,b)}\}_A\Big]Z_{AB}=0.
\end{eqnarray}

Similarly, for the transformation $B\rightarrow B+\delta B$,
where $\delta B=\dsum_{a+b=1}^{\infty}\dsum_{\check{\alpha}|\ \level(\check{\alpha})=a+b}N^{-2(a+b)}t_{\check{\alpha}}^{(a+b)}\dfrac{\partial
\mathcal{T}_{\check{\alpha}}^{(a,b)}}{\partial \bar B}$, we have
\begin{eqnarray}\label{wardB}
&& \Big[-\dsum_{a+b=1}^{\infty}\dsum_{\check{\alpha}|\ \level(\check{\alpha})=a+b}
bN^{-2(a+b)}t_{\check{\alpha}}^{(a,b)} \mathcal{T}_{\check{\alpha}}^{(a,b)}
+\dsum_{a+b=1}^{\infty}\dsum_{\check{\alpha}|\ \level(\check{\alpha})=a+b}
N^{-2(a+b)}t_{\check{\alpha}}^{(a,b)}\tilde{\Delta} \mathcal{T}_{\check{\alpha}}^{(a,b)}
\nonumber\\
&&+\dsum_{\substack{a+b=1,\\n+m=1}}^{\infty}\dsum_{\substack{\check{\alpha}|\
\level(\check{\alpha})=a+b,\\ \check{\sigma}|\ \level(\check{\sigma})=n+m}}
N^{-2(a+b+n+m)}t_{\check{\alpha}}^{(a,b)}t_{\check{\sigma}}^{(n,m)}
\{\mathcal{T}_{\check{\sigma}}^{(n,m)},\mathcal{T}_{\check{\alpha}}^{(a,b)}\}_B\Big]
Z_{AB}=0.
\end{eqnarray}

From (\ref{wardA}) and (\ref{wardB}), it is not difficult to obtain that the partition function (\ref{rainbowrg}) satisfies
\begin{eqnarray}\label{dw}
 \hat{D}Z_{AB}=\hat{W}Z_{AB},
\end{eqnarray}
where
\begin{eqnarray}\label{operatorD}
\hat{D}=\dsum_{a+b=1}^{\infty}\dsum_{\check{\alpha}|\ \level(\check{\alpha})=a+b}
aN^{-2(a+b)}t_{\check{\alpha}}^{(a,b)}
\frac{\partial}{\partial t_{\check{\alpha}}^{(a,b)}}
+\dsum_{a+b=1}^{\infty}\dsum_{\check{\alpha}|\ \level(\check{\alpha})=a+b}
bN^{-2(a+b)}t_{\check{\alpha}}^{(a,b)}\frac{\partial}{\partial t_{\check{\alpha}}^{(a,b)}},
\end{eqnarray}
\begin{eqnarray}\label{operator3}
\hat{W}&=&
\dsum_{a+b=1}^{\infty}\dsum_{\check{\alpha}|\ \level(\check{\alpha})=a+b}\sum_{k=1}^3
\sum_{\substack{\check{\beta}_1,\cdots,\check{\beta}_k,\\ a_1+\cdots+a_k+1=a,\\
b_1+\cdots+b_k=b,\\a_1\leq a_2\leq\cdots\leq a_k}}(1-\delta_{a,1})
{\Delta}_{\check{\alpha},\check{\beta}_1,\cdots,\check{\beta}_k}^{(a,b),(a _1,b_1),\cdots,(a_k,b_k)}
N^{-2(a+b)}t_{\check{\alpha}}^{(a,b)}\frac{\partial}{\partial t_{\check{\beta}_1}^{(a_1,b_1)}}
\cdots\frac{\partial}{\partial t_{\check{\beta}_k}^{(a_k,b_k)}}\nonumber\\
&+&
\dsum_{a+b=1}^{\infty}\dsum_{\check{\alpha}|\ \level(\check{\alpha})=a+b}\sum_{k=1}^3
\sum_{\substack{\check{\gamma}_1,\cdots,\check{\gamma}_k,\\ a_1+\cdots+a_k=a,\\
b_1+\cdots+b_k+1=b,\\b_1\leq b_2\leq\cdots\leq b_k}}(1-\delta_{b,1})
{\tilde{\Delta}}_{\check{\alpha},\check{\gamma}_1,\cdots,\check{\gamma}_k}^{(a,b),(a _1,b_1),\cdots,(a_k,b_k)}
N^{-2(a+b)}t_{\check{\alpha}}^{(a,b)}\frac{\partial}{\partial t_{\check{\gamma}_1}^{(a_1,b_1)}}
\cdots\frac{\partial}{\partial t_{\check{\gamma}_k}^{(a_k,b_k)}}\nonumber\\
&+&
\dsum_{\substack{a+b=1,\\n+m=1}}^{\infty}
\dsum_{\substack{\check{\alpha}|\ \level(\check{\alpha})=a+b,
\\ \check{\sigma}|\ \level(\check{\sigma})=n+m, \\ \check{\beta}|\ \level(\check{\beta})=n+m+a+b-1}}
{\Lambda}_{\check{\sigma},\check{\alpha},\check{\beta}}^{(n,m),(a,b),(n+a-1,m+b)}
N^{-2(a+b)}N^{-2(n+m)}t_{\check{\alpha}}^{(a,b)} t_{\check{\sigma}}^{(n,m)}
\frac{\partial}{\partial t_{\check{\beta}}^{(n+a-1,m+b)}}
\nonumber\\
&+&
\dsum_{\substack{a+b=1,\\n+m=1}}^{\infty}
\dsum_{\substack{\check{\alpha}|\ \level(\check{\alpha})=a+b,
\\ \check{\sigma}|\ \level(\check{\sigma})=n+m, \\ \check{\beta}|\ \level(\check{\beta})=n+m+a+b-1}}
{\tilde{\Lambda}}_{\check{\sigma},\check{\alpha},\check{\gamma}}^{(n,m),(a,b),(n+a-1,m+b)}
N^{-2(a+b)}N^{-2(n+m)}t_{\check{\alpha}}^{(a,b)} t_{\check{\sigma}}^{(n,m)}
\frac{\partial}{\partial t_{\check{\gamma}}^{(n+a-1,m+b)}}
\nonumber\\
&+&
N^{-2}t_{ (id, id, id) }^{(1,0)}\mathcal{N}_3
+N^{-2}t_{(id, id, id)}^{(0,1)}\mathcal{N}_3,
\end{eqnarray}
where $\check{\alpha}$, $\check{\beta}$ and $\check{\sigma}$ are taken from indices of connected
operators in the ring, $\mathcal{N}_3= {N_{1}} {N_{2}}  {N_{3}}$.

Let us rewrite (\ref{rainbowrg}) as
\begin{eqnarray}\label{rainbowrg2}
Z_{AB}=\sum_{s=0}^{\infty}Z_{AB}^{(s)},
\end{eqnarray}
where
\begin{eqnarray}
Z_{AB}^{(s)}&=&\int dA d\bar A dB d\bar B
\exp(- \Tr A\bar A- \Tr B\bar B)\cdot\sum_{l=0}^{\infty}
\sum_{\substack{\check{\sigma}_i|\ \level(\check{\sigma}_i)=n_i+m_i,\\n_1+m_1+\cdots+n_l+m_l=s}}
\frac{1}{l!}\langle \prod_{i=1}^{l}\mathcal{T}_{\check{\sigma}_i}^{(n_i,m_i)}\rangle
\nonumber\\&&\cdot
N^{-2(n_1+m_1+\cdots+n_l+m_l)}\prod_{i=1}^{l}t_{\check{\sigma}_i}^{(n_i,m_i)},
\end{eqnarray}
and the correlators~$\langle \prod_{i=1}^{l}\mathcal{T}_{\check{\sigma}_i}^{(n_i,m_i)}\rangle$
are defined by
\begin{eqnarray}
\langle \prod_{i=1}^{l}\mathcal{T}_{\check{\sigma}_i}^{(n_i,m_i)}\rangle
=\frac{\int dA d\bar A dB d\bar B
\mathcal{T}_{\check{\sigma}_1}^{(n_1,m_1)}\mathcal{T}_{\check{\sigma}_2}^{(n_2,m_2)}
\cdots\mathcal{T}_{\check{\sigma}_l}^{(n_l,m_l)} \exp(- \Tr A\bar A- \Tr B\bar B)}
{\int dA d\bar A dB d\bar B \exp(- \Tr A\bar A- \Tr B\bar B)}.
\end{eqnarray}

From the operators $\hat D$ and $\hat{W}$ acting on $Z_{AB}^{(s)}$,
\begin{eqnarray}\label{dzs}
\hat {D} Z_{AB}^{(s)}=sZ_{AB}^{(s)},
\end{eqnarray}
\begin{eqnarray}\label{increa}
\hat{W}Z_{AB}^{(s)}=(s+1)Z_{AB}^{(s+1)},
\end{eqnarray}
we see that the operators $\hat D$ and $\hat{W}$
are indeed the operators preserving and increasing the grading, respectively.
Thus the partition function can be realized by the $W$-representation
\begin{eqnarray}\label{exp}
Z_{AB}=\exp(\hat{W})\cdot 1.
\end{eqnarray}

Let us write the $k$-th power of the operator $\hat{W}$ as
\begin{eqnarray}\label{wm}
\hat{W}^k&=&
\sum_{i=1}^{k}\sum_{a_1+b_1+\cdots+a_i+b_i=k}\sum_{\check{\alpha}_i|\ \level(\check{\alpha}_i)=a_i+b_i}
{P}_{~\check{\alpha}_1,\cdots,\check{\alpha}_i} ^{(a_1,b_1),\cdots,(a_i,b_i)}
N^{-2k}t_{\check{\alpha}_1}^{(a_1,b_1)}\cdots t_{\check{\alpha}_{i}}^{(a_i,b_i)}\nonumber\\
&+&\sum_{i=1}^{2k}\sum_{j=1}^{3k}\sum_{\substack{a_1+b_1\cdots+a_i+b_i\\
-n_1-m_1-\cdots-n_j-m_j=k}}
\sum_{\substack{\check{\alpha}_i|\ \level(\check{\alpha}_i)=a_i+b_i,\\
\check{\beta}_j|\ \level(\check{\beta}_j)=n_j+m_j}}
{P}_{\check{\alpha}_1,\cdots,\check{\alpha}_i;(a_1,b_1),\cdots,(a_i,b_i)}
^{\check{ \beta}_1,\cdots,\check{\beta}_j;(n_1,m_1),,\cdots,(n_j,m_j)}\nonumber\\
&&\cdot N^{-2k}
t_{\check{\alpha}_1}^{(a_1,b_1)}\cdots t_{\check{\alpha}_{i}}^{(a_i,b_i)}
\frac{\partial}{\partial {t_{\check{\beta}_1}^{(n_1,m_1)}}}
\cdots \frac{\partial}{\partial {t_{\check{\beta}_{j}}^{(n_j,m_j)}}},
\end{eqnarray}
where the coefficients  ${P}_{~\check{\alpha}_1,\cdots,\check{\alpha}_i} ^{(a_1,b_1),\cdots,(a_i,b_i)}$  and
${P}_{\check{\alpha}_1,\cdots,\check{\alpha}_i;(a_1,b_1),\cdots,(a_i,b_i)}^{ \check{\beta}_1,\cdots,\check{\beta}_j;(n_1,m_1),,\cdots,(n_j,m_j)}$
are polynomials of $ {N_{1}}$, $ {N_{2}}$ and $ {N_{3}}$.

Then it is not difficult to derive the compact expression of correlators from (\ref{exp})
\begin{eqnarray}\label{corrf}
\left\langle \mathcal{T}_{\check{\alpha}_1}^{(a_1,b_1)}\cdots \mathcal{T}_{\check{\alpha}_{i}}^{(a_i,b_i)} \right\rangle
=\frac{i!}{k!\lambda_{(\check{\alpha}_1,\cdots,\check{\alpha}_i)}}
\sum_{\tau}{P}_{~\tau(\check{\alpha}_1),\cdots,\tau(\check{\alpha}_i)}^{(a_1,b_1),\cdots,(a_i,b_i)},
\end{eqnarray}
where~$k=a_1+b_1+\cdots+a_i+b_i$, $\tau$ denotes all distinct permutations of $(\check{\alpha}_1,\cdots,\check{\alpha}_i)$
and~$\lambda_{(\check{\alpha}_1,\cdots,\check{\alpha}_i)}$ is the number of $\tau$ with respect to~$\check{\alpha}_1,\cdots,\check{\alpha}_i$.

We list some correlators in (\ref{corrf}) as follows:
\begin{eqnarray}\label{correlators}
 &&\left\langle \mathcal{T}_{(id, id, id)}^{(1,0)}\right\rangle
=\left\langle \mathcal{T}_{(id, id, id)}^{(0,1)}\right\rangle
=\mathcal{N}_3,
\nonumber\\
 &&\left\langle \mathcal{T}_{((12), id, id)}^{(2,0)}\right\rangle
=\left\langle \mathcal{T}_{((12), id, id)}^{(0,2)}\right\rangle
=\mathcal{N}_3( {N_{1}}+ {N_{2}}  {N_{3}}),
\nonumber\\
 &&\left\langle \mathcal{T}_{(id, id, id)}^{(1,0)}\mathcal{T}_{(id, id, id)}^{(1,0)}\right\rangle
=\left\langle \mathcal{T}_{(id, id, id)}^{(0,1)}\mathcal{T}_{(id, id, id)}^{(0,1)}\right\rangle
=\mathcal{N}_3^2+\mathcal{N}_3,
\nonumber\\
 &&\left\langle \mathcal{T}_{(id, id, id)}^{(1,0)}\mathcal{T}_{(id, id, id)}^{(0,1)}\right\rangle
=\mathcal{N}_3^2,\left\langle \mathcal{T}_{((12), id, id)}^{(1,1)}\right\rangle
=\mathcal{N}_3 {N_{2}}  {N_{3}},
\nonumber\\
 &&\left\langle \mathcal{T}_{(id, id, id)}^{(1,0)}\mathcal{T}_{((12), id, id)}^{(2,0)}\right\rangle
=\left\langle \mathcal{T}_{(id, id, id)}^{(0,1)}\mathcal{T}_{((12), id, id)}^{(0,2)}\right\rangle
=\mathcal{N}_3( {N_{1}}+ {N_{2}}  {N_{3}})
 +\mathcal{N}_3^2( {N_{1}}+ {N_{2}}  {N_{3}}),
\nonumber\\
 &&\left\langle \mathcal{T}_{(id, id, id)}^{(1,0)}\mathcal{T}_{((12), id, id)}^{(0,2)}\right\rangle
=\left\langle \mathcal{T}_{(id, id, id)}^{(0,1)}\mathcal{T}_{((12), id, id)}^{(2,0)}\right\rangle
=\mathcal{N}_3^2( {N_{1}}+ {N_{2}}  {N_{3}}),
\nonumber\\
 &&\left\langle \mathcal{T}_{((123), id, id)}^{(3,0)}\right\rangle
=\left\langle \mathcal{T}_{((123), id, id)}^{(0,3)}\right\rangle
=\mathcal{N}_3^2+\mathcal{N}_3+\mathcal{N}_3[ {N_{1}}^2
 +( {N_{2}}  {N_{3}})^2
 +2 {N_{1}} {N_{2}}  {N_{3}}],
\nonumber\\
 &&\left\langle \mathcal{T}_{(id, (123), (12))}^{(3,0)}\right\rangle
=\left\langle \mathcal{T}_{(id, (123), (12))}^{(0,3)}\right\rangle
=\mathcal{N}_3^2 {N_{3}}
+\mathcal{N}_3( {N_{2}}^2 {N_{3}}
+ {N_{1}}^2 {N_{3}}
+2 {N_{1}} {N_{2}}+ {N_{3}}),
\nonumber\\
 &&\left\langle \mathcal{T}_{(id, (123), (132))}^{(3,0)}\right\rangle
=\left\langle \mathcal{T}_{(id, (123), (132))}^{(0,3)}\right\rangle
=3\mathcal{N}_3^2
+\mathcal{N}_3( {N_{1}}^2+ {N_{2}}^2+ {N_{3}}^2),
\nonumber\\
 &&\left\langle \mathcal{T}_{((123), id, id)}^{(2,1)}\right\rangle
=\left\langle \mathcal{T}_{((123), id, id)}^{(1,2)}\right\rangle
=\mathcal{N}_3^2+\mathcal{N}_3( {N_{2}} {N_{3}})^2,
\nonumber\\
 &&\left\langle \mathcal{T}_{(id, (123), (12))}^{(2,1)}\right\rangle
=\left\langle \mathcal{T}_{(id, (123), (12))}^{(1,2)}\right\rangle
=\mathcal{N}_3^2 {N_{3}}
 +\mathcal{N}_3( {N_{2}}^2 {N_{3}}
 +2 {N_{1}} {N_{2}}+ {N_{3}}),
\nonumber\\
 &&\left\langle \mathcal{T}_{(id, id, id)}^{(1,0)} \mathcal{T}_{(id, id, id)}^{(1,0)}
\mathcal{T}_{(id, id, id)}^{(0,1)}\right\rangle
=\left\langle \mathcal{T}_{(id, id, id)}^{(1,0)} \mathcal{T}_{(id, id, id)}^{(0,1)}
\mathcal{T}_{(id, id, id)}^{(0,1)}\right\rangle
=\mathcal{N}_3^3+\mathcal{N}_3^2,
\nonumber\\
 &&\left\langle \mathcal{T}_{((12), id, id)}^{(1,1)} \mathcal{T}_{(id, id, id)}^{(1,0)}\right\rangle
=\left\langle \mathcal{T}_{((12), id, id)}^{(1,1)} \mathcal{T}_{(id, id, id)}^{(0,1)}\right\rangle
=\mathcal{N}_3^2 {N_{2}} {N_{3}}+\mathcal{N}_3
 {N_{2}} {N_{3}},
\nonumber\\
 &&\left\langle (\mathcal{T}_{(id, id, id)}^{(1,0)} )^{i}\right\rangle
=\left\langle (\mathcal{T}_{(id, id, id)}^{(0,1)} )^{i} \right\rangle
=\prod_{j=0}^{i-1}(\mathcal{N}_3+j).
 \end{eqnarray}

\subsection{Perturbative collective field theory}

Let us consider the Hamiltonian \cite{Mahu2020}
\begin{eqnarray}\label{Hamilton}
H=-\frac{\partial}{\partial A_i^{j ,k}}\frac{\partial}{\partial \bar {A}_i^{j ,k}}
-\frac{\partial}{\partial B_i^{j ,k}}\frac{\partial}{\partial \bar {B}_i^{j ,k}}
+ \frac{1}{4}A_i^{j ,k}\bar {A}_i^{j ,k}+\frac{1}{4}B_i^{j ,k}\bar {B}_i^{j ,k}.
\end{eqnarray}

We introduce the collective variables
\begin{eqnarray}
\Phi_k=\Tr(e^{ikT}),
\end{eqnarray}
where $T$ is a matrix on the vector space with elements
\begin{eqnarray} \label{Tterms} T_{ {i_{1}}}^{ {i_{2}}}=
A_{ {i_{1}}}^{{ {j_1}},{ {j_2}}}
\bar{A}^{ {i_{2}}}_{{ {j_1}},{ {j_2}}}+
B_{ {i_{1}}}^{{ {j_1}},{ {j_2}}}
\bar{B}^{ {i_{2}}}_{{ {j_1}},{ {j_2}}}.
\end{eqnarray}
Note that it is different from the discussion in \cite{Mahu2020} where the second term in (\ref{Tterms})
was dropped.

Let us introduce the field $\Phi(x)=\int\frac{dk}{2\pi}e^{-ik x}\Phi_k,$
which is the eigenvalues density of matrix $T$ \cite{Tribelhorn}.
Using the collective valiables $\Phi_k$,
we rewrite the kinetic terms in $(\ref{Hamilton})$ as
\begin{eqnarray}
-\frac{\partial}{\partial A_i^{j ,k}}\frac{\partial}{\partial \bar {A}_i^{j ,k}}
-\frac{\partial}{\partial B_i^{j ,k}}\frac{\partial}{\partial \bar {B}_i^{j ,k}}
&=&-T_l^i\frac{\partial}{\partial T_l^{j }}\frac{\partial}{\partial T_i^{j }}
-2 {N_2} {N_3}\frac{\partial}{\partial T_i^{i}}\nonumber\\
&=&\int dk\int dk^{'}\Omega_{k,k^{'}}\pi_k\pi_{k^{'}}+\int dk \omega_k\pi_k,
\end{eqnarray}
where
$\pi_k=\frac{1}{i}\frac{\delta}{\delta\Phi_k}$, $\Omega_{k,k^{'}}$ and $\omega_k$ are given by
\begin{eqnarray}\label{omega}
&&\Omega_{k,k^{'}}=T_l^i\frac{\partial\Phi_k}{\partial T_l^{j }}\frac{\partial\Phi_{k^{'}}}{\partial T_i^{j }}
=ikk^{'}\frac{\partial}{\partial k}\Phi_{k+k^{'}},
\nonumber\\
&&\omega_k=-T_l^i\frac{\partial}{\partial T_l^{j }}(\frac{\partial\Phi_{k}}{\partial T_i^{j }})
-2 {N_2} {N_3}\frac{\partial\Phi_{k}}{\partial T_i^{i }}
=k\int_0^1d\tau \Phi_{(1-\tau )k}-2ik {N_2} {N_3}\Phi_k.
\end{eqnarray}
Then by the fourier transformations of (\ref{omega}), we have
\begin{eqnarray}
&&\Omega(x,x^{'})=\partial_x\partial_x^{' }(x\Phi(x)\delta(x-x^{'}),
\nonumber\\
&&\omega(x)=2\partial_x\fint dy\Phi(x)\Phi(y)\frac{x}{x-y}+[2 {N_2} {N_3}
- {N_1}]\partial_x\Phi(x).
\end{eqnarray}

We may further write the Hamiltonian in terms of the collective variables.
By performing similarity transformation on Hamiltonian to hermitian it,
it gives an equality
\begin{eqnarray}\label{omegaequ}
2\int dx^{'}\Omega(x,x^{'},\Phi)C(x^{'})+\omega(x,\Phi)=0.
\end{eqnarray}
From $(\ref{omegaequ})$, we have
\begin{eqnarray}\label{partial_xC}
\partial_x C(x)=\frac{1}{2x}\cdot\fint dy\frac{2x\Phi(y)}{x-y}+\frac{(2\cdot\frac{ {N_2} {N_3}}{ {N_1}}-1)
 {N_1}}{2x}.
\end{eqnarray}

The desired Hamiltonian can be finally written as
\begin{eqnarray}\label{Hamiltonian}
H=\int dx\frac{\partial \pi(x)}{\partial x}\phi(x)\frac{\partial \pi(x)}{\partial x}+V_{eff},
\end{eqnarray}
where $\pi(x)=\frac{1}{i}\frac{\delta}{\delta \Phi(x)}$, the effective potential is given by
\begin{eqnarray}
&&V_{eff}=\int dx[\frac{\pi^{2}}{3}x{\Phi(x)}^3
+\frac{(2\cdot\frac{ {N_2} {N_3}}{ {N_1}}-1)^2}{4x}
 {N_1}^2\Phi(x)+\frac{x}{4}\Phi(x)-\Lambda\Phi(x)],
\end{eqnarray}
$\Lambda$ is the lagrange multiplier enforces the constraint $\int dx \Phi(x)= {N_1}$.

The classical field should minimize the effective potential. By $\frac{\delta V_{eff}}{\delta\Phi}=0$,
we have
\begin{eqnarray}\label{effectiveeq}
\pi^2x\Phi^2
+\frac{(2\cdot\frac{ {N_2} {N_3}}{ {N_1}}-1)^2}{4x}
 {N_1}^2
+\frac{x}{4}-\Lambda=0.
\end{eqnarray}

By requiring $2\Lambda-\sqrt{4\Lambda^2- {N_1}^2}\leq x \leq2\Lambda+\sqrt{4\Lambda^2- {N_1}^2}$
and taking the limit $\frac{ {N_2} {N_3}}{ {N_1}}\rightarrow1$ and the multiplier
$\Lambda=\frac{3}{2} {N_1}$ in (\ref{effectiveeq}), we obtain the classical collective field
\begin{eqnarray}
\Phi(x)=\frac{1}{2\pi}\sqrt{\frac{6 {N_1}}{x}-1-\frac{ {N_1}^2}{x^2}}.
\end{eqnarray}

Then the collective field computation gives
\begin{eqnarray}\label{collresult}
\int^{(3+2\sqrt{2}) {N_1}}_{(3-2\sqrt{2}) {N_1}} dx \Phi(x)x^n
=C_n {N_1}^{n+1},
\end{eqnarray}
where
\begin{eqnarray}
&&C_n
=4\cdot3^{n-1}
-6(1+(-1)^{n})\frac{(2\sqrt{2})^{n-2}(n-1)(n-1)!!}{n!!}
-2(1+(-1)^{n+1})\frac{(2\sqrt{2})^{n-1}n!!}{(n+1)!!}
\nonumber\\
&&+2\frac{(n-1)!!}{n!!}\sum_{k=2}^{n-1}
(2\sqrt{2})^{k-2}3^{n-k-1}(1+(-1)^{k})
[8\binom{n-1}{k}-9\binom{n-1}{k-2}].
\end{eqnarray}

The first several ones of (\ref{collresult}) are
\begin{eqnarray}\label{integralsolution}
\int_{(3-2\sqrt{2}) {N_1}}^{(3+2\sqrt{2}) {N_1}}dx\Phi(x)x &=&2 {N_1}^2 ,\nonumber\\
\int_{(3-2\sqrt{2}) {N_1}}^{(3+2\sqrt{2}) {N_1}}dx\Phi(x)x^2 &=&6 {N_1}^3,\nonumber\\
\int_{(3-2\sqrt{2}) {N_1}}^{(3+2\sqrt{2}) {N_1}}dx\Phi(x)x^3 &=&22 {N_1}^4 .
\end{eqnarray}

As done in Ref. \cite{Tribelhorn}, let us consider the limit that $N_i\rightarrow\infty,~(i=1,2,3)$ and setting
$ {N_1}= {N_2} {N_3}$ in the correlators $\left\langle \Tr T^n\right\rangle$,
then the classical collective solution (\ref{collresult}) gives the highest power term of $ {N_1}$ in $\left\langle \Tr T^n\right\rangle$.
Thus in this large $N_i$ limit, we write the correlators as
\begin{eqnarray}\label{Tn}
\left\langle \Tr T^n\right\rangle
=\left\langle \Tr (A\bar A+B\bar B)^n\right\rangle
=C_n {N_1}^{n+1}+\circ( {N_1}^n),
\end{eqnarray}
where the expression of $\Tr (A\bar A+B\bar B)^n$ contains all the tree operators
$\mathcal{T}^{(p,n-p)}_{(\sigma, id , id)}$, $\sigma\in S_n,$ in $\tilde{S}$,
$\tilde{S}$ is a closed ring generated by $\mathcal{T}_{((12), id , id)}^{(2,0)}$,
$\mathcal{T}_{((12), id , id)}^{(0,2)}$ and $\mathcal{T}_{((12), id , id)}^{(1,1)}$
through addition, multiplication, cut and join operations.

Let us list the correlators corresponding to (\ref{integralsolution})
\begin{eqnarray}\label{N1=N2N3corre}
\includegraphics[width=0.75\textwidth]{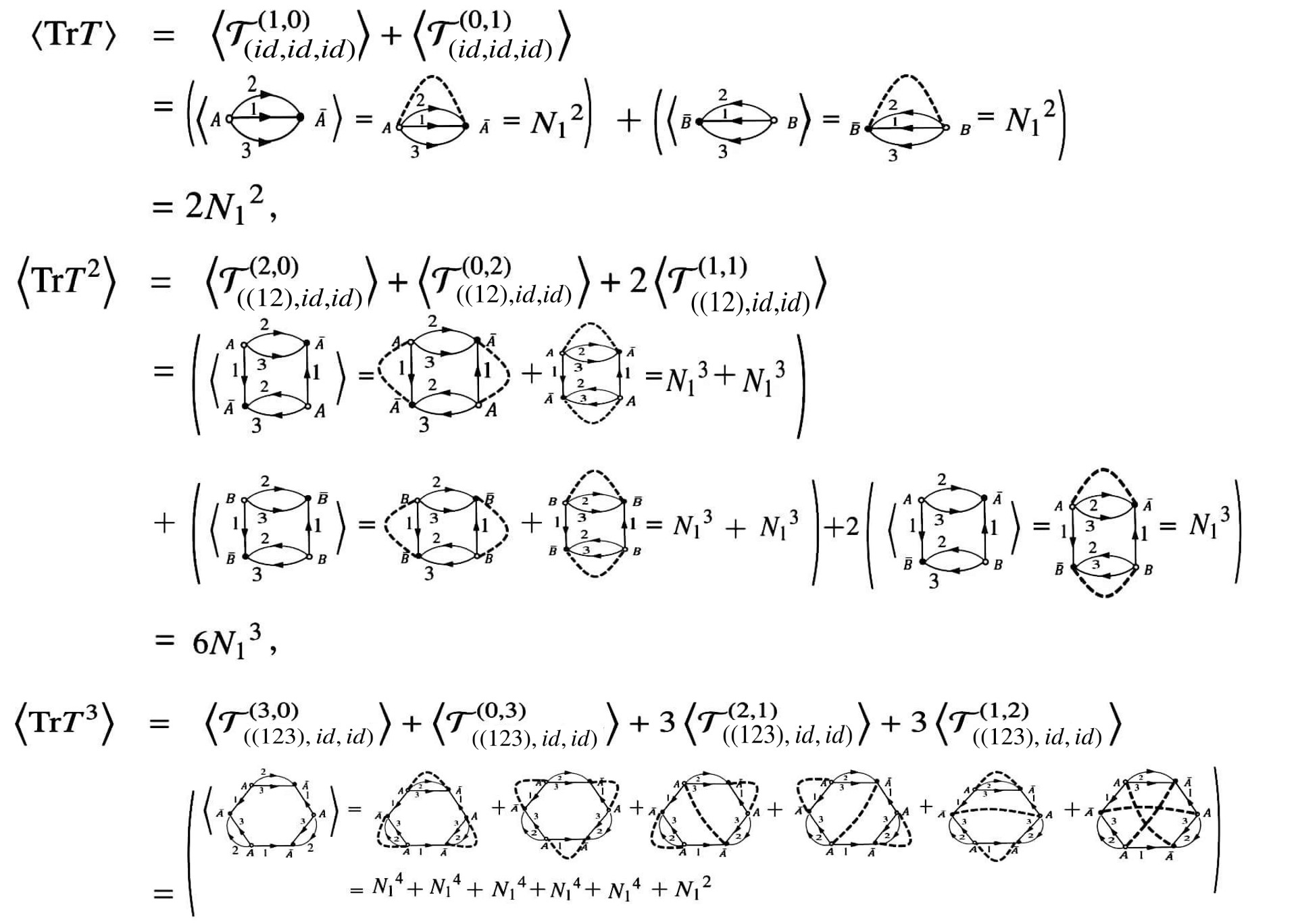}\nonumber\\
\includegraphics[width=0.76\textwidth]{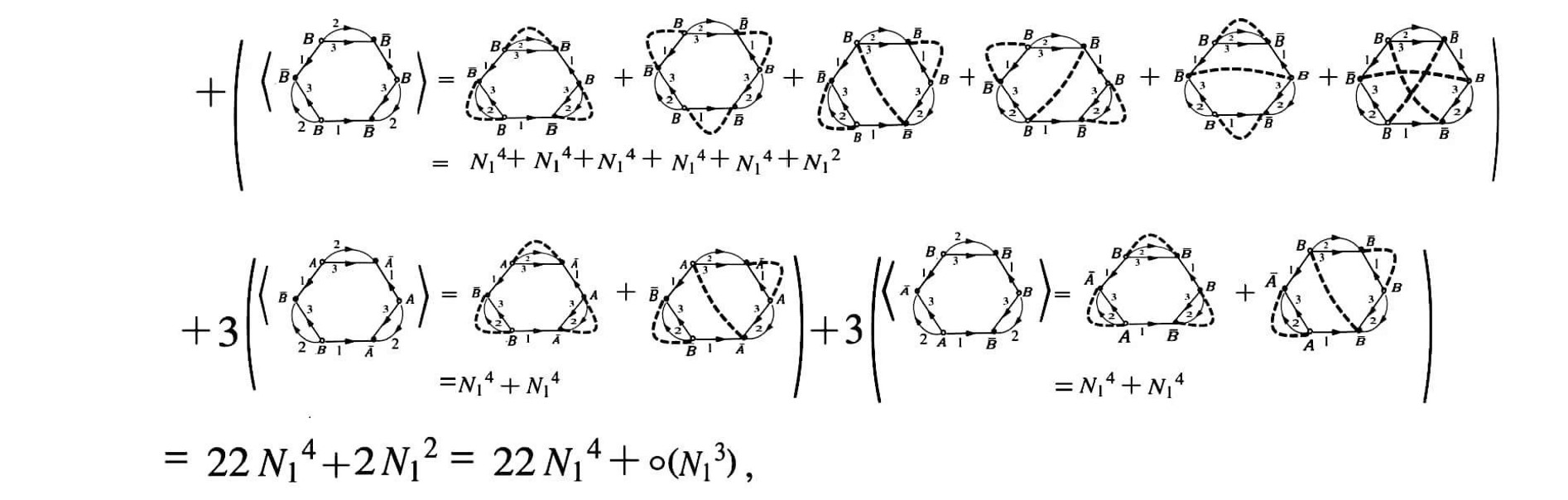}
\end{eqnarray}
where the dotted lines depict the Feynman propagators.

\begin{figure}[H]
\centering
\includegraphics[width=0.5\textwidth]{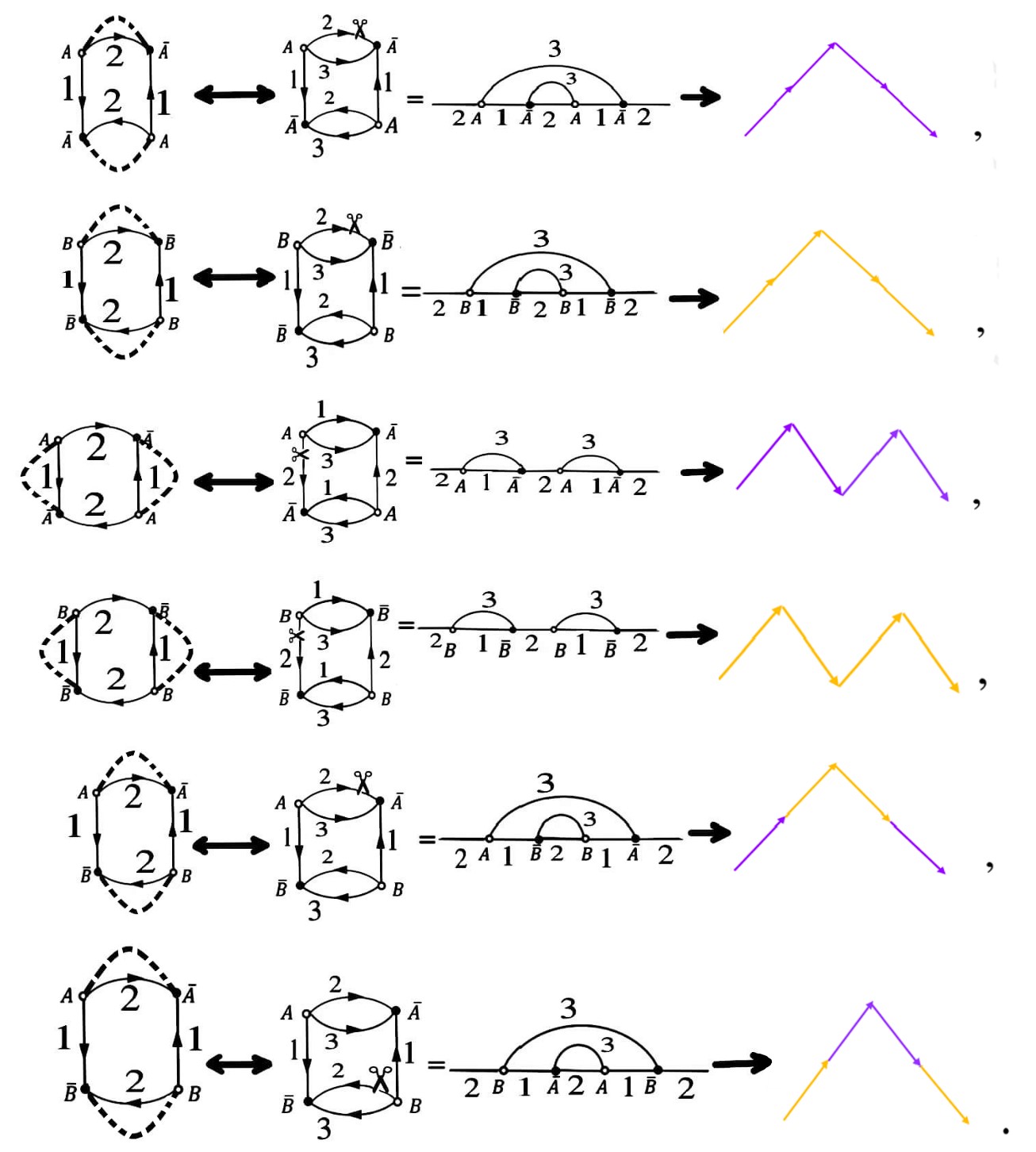}
\caption{Feynman diagrams contributing to highest power terms of~$ {N_1}$ in~$\langle \Tr T^2\rangle$
and the corresponding length-$4$ two colored Dyck walks.}
\label{T-spinchain-2}
\end{figure}

The Feynman diagrams of operators in $\tilde{S}$ are circles \textbf{I} (see Figs. \ref{loopr=3}(a)).
It implies that the lines of color $2$ and $3$ always simultaneous appear in these Feynman diagrams.
We delete the lines index $3$ in these Feynman diagrams (see examples in the left part of Fig. \ref{T-spinchain-2}),
which do not affect the results in this section. Thus the Wick contractions contributing to the highest power terms of~$ {N_1}$
in~(\ref{N1=N2N3corre}) can be depicted as circles  \textbf{I}  and \textbf{II} with disjoint dotted lines connecting~$A$ and~$\bar{A}$,
or~$B$ and~$\bar{B}$, where dotted lines denote Wick contractions. Then by changing the dotted lines into  lines of color $3$,
we obtain tree operators in the closed ring~$\check{S}$ generated by five keystones
$\mathcal{T}_{((12), id , id)}^{(2,0)}$, $\mathcal{T}_{((12), id , id)}^{(0,2)}$,
$\mathcal{T}_{(id, (12) , id)}^{(2,0)}$, $\mathcal{T}_{(id, (12), id)}^{(0,2)}$
and $\mathcal{T}_{((12), id , id)}^{(1,1)}$ (see examples in the second part on the left of Fig.\ref{T-spinchain-2}).

For the number of the length-$2n$ two colored Dyck walks, it is counted by $2^n$ multiples of the~$n$-th Catalan number (\ref{Catalan}).
Based on the correspondence between Fredkin spin chain and tree operators in the previous section,
we see that the number of spin chains corresponding to tree operators with level-$n$ in $\check{S}$ is given by $C_n$.
In other words, $C_n$ counts the number of special length-$2n$ two colored Dyck walks which satisfy the following two conditions:
(i) The colors of the first and  last steps are same.
(ii) The colors of~$2k$-th and  $(2k+1)$-th steps are same, but
for the $(2k-1)$-th and $2k$-th steps, $k=1,2,\cdots,n-1$, their colors don't have to be same.
In Fig. \ref{T-spinchain-2}, we take $\langle \Tr T^2\rangle$ in (\ref{N1=N2N3corre}) as an example
to draw the corresponding six spin chains.

\subsection{Free energy and large $N$ limit}
Let us consider the free energy $\mathcal{F}$ of the two-tensor model (\ref{rainbowrg})
\begin{eqnarray} \label{FE}
\mathcal{F}&=&-\ln Z_{AB}\nonumber\\
&=&
\sum_{s,l=0}^{\infty}
\sum_{\substack{\lambda\mapsto l\ , length(\lambda)=p,
\\ \check{\alpha}_{i}|\ \level ( \check{\alpha}_{i})=n_i+m_i,
\\n_1+m_1+\cdots+n_{\lambda_p}+m_{\lambda_p}=s}}
\frac{(-1)^{p+1}}{l!p}S(\mu)\tilde{S}(\lambda)
\langle\mathcal{T}_{\check{\alpha}{_1}}^{(n_1,m_1)}\cdots
\mathcal{T}_{\check{\alpha}_{{\lambda_1}}}^{(n_{\lambda_1},m_{\lambda_1})}\rangle
\langle\mathcal{T}_{\check{\alpha}_{{\lambda_1+1}}}^{(n_{\lambda_1+1},m_{\lambda_1+1})}
\cdots\mathcal{T}_{\check{\alpha}_{{\lambda_2}}}^{(n_{\lambda_2},m_{\lambda_2})}\rangle
\nonumber\\
&&\cdot
\cdots
\langle\mathcal{T}_{\check{\alpha}_{{\lambda_{p-1}+1}}}^{(n_{\lambda_{p-1}+1},m_{\lambda_{p-1}+1})}
\cdots\mathcal{T}_{\check{\alpha}_{{\lambda_p}}}^{(n_{\lambda_p},m_{\lambda_p})}\rangle
\prod_{i=1}^{l}t_{\check{\alpha}_i}^{(n_i,m_i)}N^{-2(n_1+m_1+\cdots+n_l+m_l)}
,
\end{eqnarray}
where the partitions $\lambda$ and $\mu$ are $\lambda=(1^{n_1+m_1},2^{n_2+m_2},\cdots,p^{n_p+m_p})$
and $ \mu=(1^{u_1+v_1},2^{u_2+v_2},\cdots,p^{u_p+v_p})$,
$S(\mu)=\binom{\sum_i n_i+m_i}{u_1+v_1 }\binom{\sum_i n_i+m_i -(u_1+v_1)}{u_2+v_2 }
\cdots 1,$
$\tilde{S}(\lambda)=(n_1+m_1)!(n_2+m_2)!\cdots(n_p+m_p)!$.

Here we have taken $ {N_{1}}= {N_{2}}= {N_{3}}=N$ in (\ref{FE}),
then the correlators can be written as
\begin{eqnarray}
\langle\mathcal{T}_{\check{\alpha}_1}^{(n_1,m_1)}\cdots\mathcal{T}_{\check{\alpha}_l}^{(n_l,m_l)}\rangle
&=&N^{2(n_1+m_1)+1+\cdots+2(n_l+m_l)+1}\cdot \frac{l!c_l}{(n_1+m_1+\cdots+n_l+m_l)!
\lambda_{\check{\alpha}_1,\cdots,\check{\alpha}_l}}
\nonumber\\
&&+\circ(N^{2(n_1+m_1)+1+\cdots+2(n_l+m_l)+1}),
\end{eqnarray}
where $c_l$ is a constant.

Thus in large $N$ limit, we have
\begin{eqnarray}\label{F2}
\mathcal{F}&\sim&
\sum_{\check{\alpha}_1|\ \level (\check{\alpha}_1)=n_1+m_1=0}^{\infty}
\frac{N}{(n_1+m_1)!} t_{\check{\alpha}_1}^{(n_1,m_1)}.
\end{eqnarray}

For the degrees of Feynman graphs~$\omega(\mathcal{G})$
and gauge-invariant operators~$\omega(\mathcal{T})$ in the $D$-colored tensor model,
there are the relations as follows \cite{GurauRyan2012}:
\begin{eqnarray}\label{FGFT}
|F_{\mathcal{G}}|&=&\frac{D(D-1)}{2}v+D-\frac{2}{(D-1)!}\omega(\mathcal{G}),\nonumber\\
|F_{\mathcal{T}}|&=&\frac{(D-1)(D-2)}{2}v+D-1-\frac{2}{(D-2)!}\omega(\mathcal{T}),
\end{eqnarray}
where $|F_{\mathcal{G}}|$ and $|F_{\mathcal{T}}|$ denote the number of faces in the Feynman diagrams and
the gauge-invariant operators respectively, and $2v$ denotes the number of vertices in the Feynman diagrams.

Then we have the number of faces obtained only by the Wick contractions, i.e.,
\begin{eqnarray}\label{deg}
|F_{\mathcal{G}}|-|F_{\mathcal{T}}|=2v+1-\omega(\mathcal{G})+2\omega(\mathcal{T}),
\end{eqnarray}
where we have taken $D=3$ in $|F_{\mathcal{G}}|$ and $|F_{\mathcal{T}}|$.

\begin{figure}[H]
\centering
\includegraphics[width=0.7\textwidth]{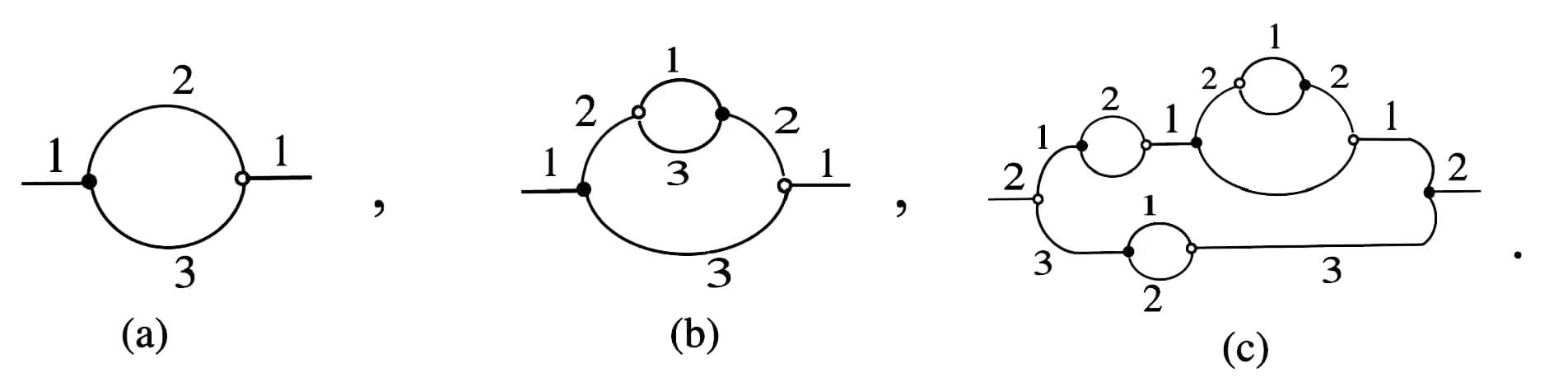}
\caption{ Examples of melonic graphs.}
\label{melonic}
\end{figure}

From (\ref{F2}) and (\ref{deg}), there is a  relation between the free energy in large $N$ limit
and melonic graphs as follows.
The highest power of~$N$ in each Gaussian average of the gauge-invariant operator is~$2v+1$.
It is given by the Feynman diagram with degree zero i.e., melonic graphs.
The melonic graphs can be obtained from a elementary melon (see Fig. \ref{melonic}(a)) with fewer vertices by
inserting two vertices connected by $2$ edges on one of the edge of the elementary melon which also
be called the dressed melon (see Figs. \ref{melonic}(b) and \ref{melonic}(c)).

\section{Conclusions}

We have given the keystone operators which are the gauge-invariant operators. Then by the keystone operators with addition,
multiplication, cut and join operations, we constructed a graded ring with tree and loop operators and enumerated the
operators in the graded ring. We have also taken the ring $S_2$ and $S_3$ as examples to analyze the kernel and cokernel
of the cut-and-join structure in these rings. In terms of the keystones operators, connected tree and loop operators in the
ring, we have constructed a two-tensor model with order-3. Moreover we showed that it can be realized by
acting on elementary function with exponent of the given operator. By means of the $W$-representation of this two-tensor
model, we have derived the compact expression of correlators and presented the free energy in large $N$ limit.

The Fredkin spin chain exhibits violation of the cluster decomposition property and of the area law for the entanglement entropy,
with the presence of anomalous and extremely fast propagation of the excitations after driving the system out-of-equilibrium.
It seems that this model extremely promising for applications in quantum information and communication processes.
We have established a correspondence between two colored Dyck walks in Fredkin spin chain and tree operators in tensor model.
For the number of the length-$2n$ two colored Dyck walks, it is counted by $2^n$ multiples of the~$n$-th Catalan number.
Based on the classification Dyck walks, we gave the number of tree operators with level-$l$, i.e., $\sharp_{l}^{tree}$.
Then the number of connected loop operators can be obtained from the difference between the numbers of connected operators
and tree operators. On the other hand, by the collective field computation, we gave the highest power term of $ {N_1}$
in correlators $\left\langle \Tr T^n\right\rangle$ with ${N_1}= {N_2} {N_3}$.
It was noted that the coefficient $C_n$ of the highest power term of ${N_1}$
indeed counts the number of special length-$2n$ colored Dyck walks. For the entanglement entropy of the Fredkin spin chain,
it exhibits the entanglement scaling beyond logarithmic scaling in the ordinary critical systems.
Quite interestingly, we found that this result can be easily showed from the viewpoint of tensor model.
For further research, it would be interesting to explore more properties of the tensor model
and Fredkin spin chain via such kind of correspondence.

\setcounter{equation}{0}
\renewcommand\theequation{A.\arabic{equation}}
\appendix
\section{}
We list the cut operation $\Delta$ acting on the level-$3$ gauge-invariant operators as follows:
\begin{eqnarray}\label{cutlevel}
 \centering
   \includegraphics[width=16cm,height=7cm]{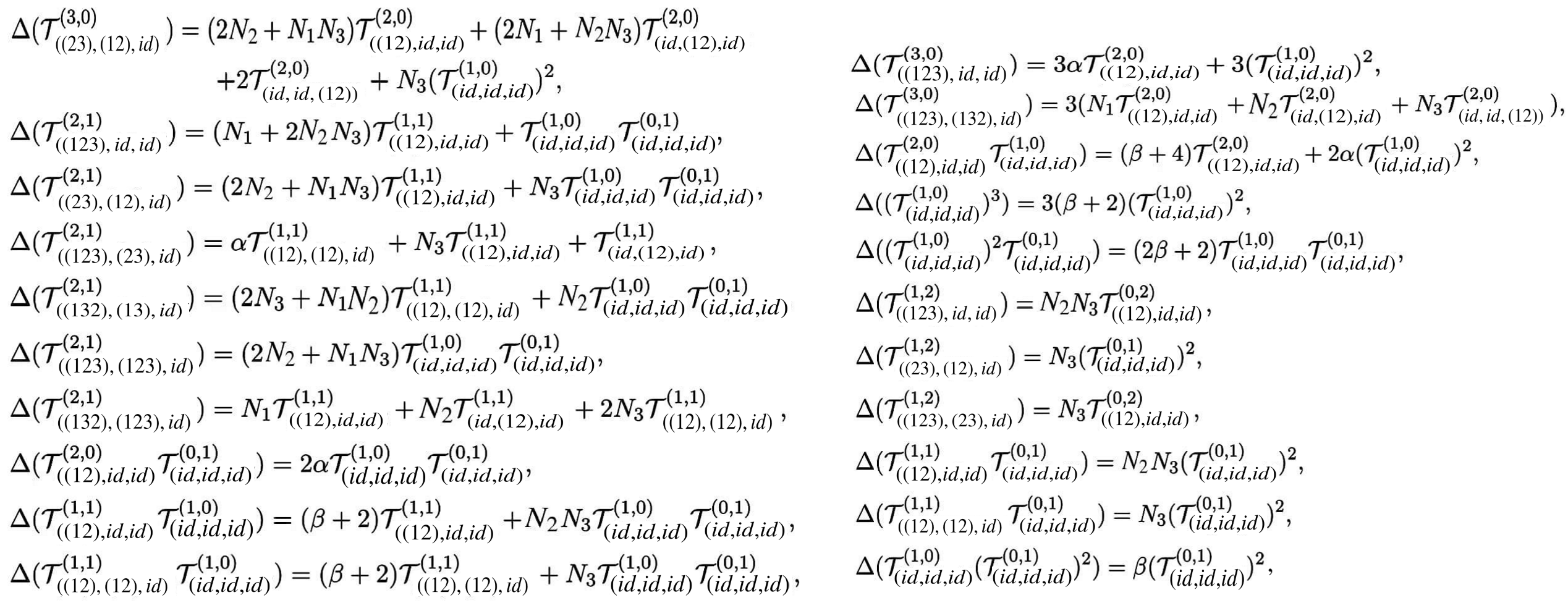}
\end{eqnarray}
then we may give the kernel
\begin{eqnarray}\label{cutkernel}
 \centering
  \includegraphics[width=15.25cm,height=11cm]{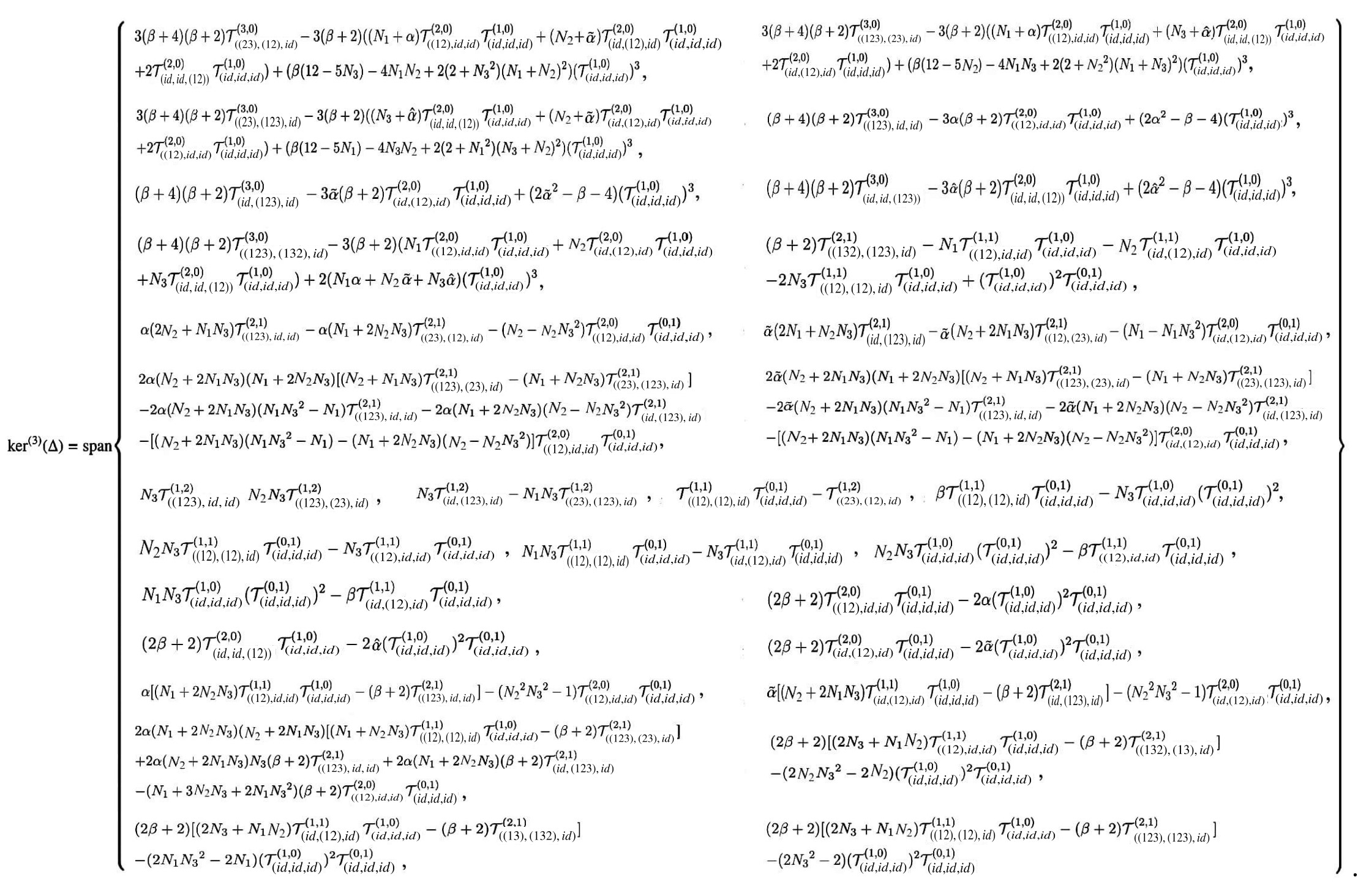}
 \end{eqnarray}

\section *{Acknowledgments}
We are grateful to J. Ben Geloun and S. Ramgoolam for helpful comments.
We are also indebted to the referee of this paper for stimulating questions.
This work is supported by the National Natural Science Foundation of China (No. 11875194).


\end{document}